%% file: dstrap_revtex.tex
\pdfoutput=1
\documentclass[aps,twocolumn,prd,preprintnumbers, groupedaddress,nofootinbib,10pt]{revtex4-1}
\usepackage{amsmath,amssymb,amsthm}
\usepackage{bm}
\usepackage[usenames,dvipsnames]{xcolor}
\usepackage{graphicx}
\usepackage{float}
\usepackage{array}
\usepackage{multirow}
\usepackage{rotfloat}
\usepackage{subcaption}
\usepackage[all]{xy}
\usepackage[normalem]{ulem}
\usepackage[hidelinks]{hyperref}
\usepackage{cleveref}
\usepackage{pgf}
\usepackage{soul}

\begin{document}

\title{Bootstrapping $(D, D)$ Conformal Matter}
\vspace*{-3cm} 
\begin{flushright}
{\tt DESY 21-165}\\
\end{flushright}

\author{Florent Baume}
\email[email: \texttt{fbaume@sas.upenn.edu}]{}
\affiliation{Department of Physics and Astronomy, University of Pennsylvania, Philadelphia, PA 19104, USA}
\author{Craig Lawrie}
\email[\texttt{gmail}: \texttt{craig.lawrie1729}]{}
\affiliation{Deutsches Elektronen-Synchrotron DESY, Notkestr.~85, 22607 Hamburg, Germany}
\affiliation{Department of Physics and Astronomy, University of Pennsylvania, Philadelphia, PA 19104, USA}

\begin{abstract}
\noindent 

We use the numerical conformal bootstrap to study six-dimensional
$\mathcal{N}=(1,0)$ superconformal field theories with flavor symmetry
$\mathfrak{so}_{4k}$. We present evidence that minimal $(D_k, D_k)$ conformal
matter saturates the unitarity bounds for arbitrary $k$. Furthermore, using the
extremal-functional method, we check that the chiral-ring relations are correctly reproduced, extract the anomalous dimensions of low-lying long superconformal multiplets, and find hints for novel OPE selection rules involving type-$\mathcal{B}$ multiplets.

\end{abstract}

\maketitle

\section{Introduction}

For the last three decades, string theory has been a potent apparatus for the
construction of quantum field theories (QFTs). The power of this approach finds its source in the way the parameters and properties of the quantum field theory are governed by geometrical and topological features of the compactification space. Conformal field theories (CFTs) are an especially interesting class of
QFTs; they exist at the fixed points of the renormalization group flows, and thus they can provide important insights into the
nature of quantum field theory.

In this work, our focus is on CFTs with additional supersymmetry, known as
superconformal field theories (SCFTs).  From a pure field theory perspective,
it was long unknown whether interacting superconformal field theories exist in
five or six dimensions, despite the existence of the appropriate superconformal
algebras  \cite{Nahm:1977tg}. In part, the challenge in constructing such
theories lies in the fact that there are no supersymmetry-preserving marginal
deformations in five or six dimensions \cite{Cordova:2016xhm}, and thus the
usual techniques of perturbation theory cannot be applied. In the 1990s, it was
discovered that compactifications of Type IIB string theory on non-compact K3
surfaces give rise to exotic six-dimensional theories, with
sixteen supercharges, whose constituent objects are tensionless strings
\cite{Witten:1995zh}, and it was soon realized that they are in fact
superconformal field theories \cite{Seiberg:1996qx}. This is a prime example of
\emph{geometric engineering}; each such 6d SCFT is associated to a finite
subgroup of $SU(2)$, $\Gamma$, as the K3 surfaces are all locally of the
orbifold form $\mathbb{C}^2/\Gamma$.

One of the recent successes of this technique of geometric engineering of
quantum field theories is the enumeration of a vast landscape of
six-dimensional superconformal field theories with eight supercharges obtained
via F-theory \cite{Heckman:2013pva, Heckman:2015bfa}.  The SCFTs realized by
this construction have a quiver-like structure;
the ``links''
appearing in these quivers are themselves non-trivial interacting SCFTs with a
$\mathfrak{g} \oplus \mathfrak{g}^\prime$ flavor algebra -- they are a
generalization of an $\mathfrak{su}_n \oplus \mathfrak{su}_m$ bifundamental
hypermultiplet -- known as minimal $(\mathfrak{g}, \mathfrak{g}^\prime)$ 
conformal matter \cite{DelZotto:2014hpa}. These conformal matter theories have
a simple construction from the perspective of M-theory: for each ADE algebra
$\mathfrak{g}^\prime = \mathfrak{g}$, the minimal conformal matter theory lives
on the worldvolume of a single M5-brane probing a $\mathbb{C}^2/\Gamma$
orbifold, where $\Gamma$ is the finite subgroup of $SU(2)$ of the same ADE-type
as $\mathfrak{g}$.

Despite marking a significant milestone in our understanding of six-dimensional
field theories, it remains unknown to what extent the landscape of
consistent 6d SCFTs matches that obtained from geometric constructions.  There
are six-dimensional SCFTs which are constructed with ``frozen'' 7-branes,
and these are not captured by the geometric constructions of
\cite{Heckman:2013pva, Heckman:2015bfa}, however such SCFTs may still be
obtainable from an F-theory origin \cite{Tachikawa:2015wka,Bhardwaj:2018jgp}.
Similarly, there are putative 6d SCFTs that appear non-anomalous from the
bottom-up, field-theoretic perspective, however there is no known way to
engineer these theories from F-theory; such theories are widely believed to be
inconsistent for a variety of indirect reasons  \cite{Ohmori:2015pia,Morrison:2016djb}, however this has not been
rigorously established. As a final
example, there is a seemingly consistent spectrum with $\mathfrak{su}_3$ flavor
symmetry, for which the Higgs branch is the one-instanton moduli space of
$SU(3)$ \cite{Shimizu:2017kzs}; this theory does not have a known geometric
construction.  It is an open question, which can be considered as part of the
swampland program, whether all consistent 6d SCFTs have an origin in string
theory.

The main drawback of the geometric engineering approach is that, up to a small
collection of protected quantities such as the central charges, it is unknown how to
access the conformal data, namely the conformal dimensions and OPE
coefficients, from the geometry. This raises the question of whether
other techniques, when used in conjunction with geometry, can be employed to
learn about those conformal quantities. Some work towards the determination of
the conformal dimensions for certain classes of unprotected operators in 6d
$(1,0)$ SCFTs was initiated in \cite{Heckman:2014qba,Baume:2020ure}. 

Concurrently, but transversely, to the development of the geometric program,
the conformal bootstrap \cite{Ferrara:1973yt,Polyakov:1974gs} has known renewed
interest, sparked in \cite{Rattazzi:2008pe}, and provides a technique for
bounding the conformal data of a CFT. Using the associativity of the operator
product expansion (OPE), the ethos of the bootstrap is to harness the power of
unitarity to impose strict bounds on the values that the conformal data of a(n
S)CFT can take, with minimal assumptions on the spectrum of the theory.
The superconformal bootstrap has been applied to SCFTs with at least eight
supercharges; in three
\cite{Chester:2014fya,Liendo:2015cgi,Agmon:2017xes,Agmon:2019imm,Chang:2019dzt,Binder:2020ckj}
and four
\cite{Beem:2013qxa,Beem:2014zpa,Alday:2014qfa,Lemos:2015awa,Liendo:2015ofa,Beem:2016wfs,Lemos:2016xke,Liendo:2016ymz,Ramirez:2016lyk,Cornagliotto:2017snu,Liendo:2018ukf,Gimenez-Grau:2019hez,Bissi:2020jve,Gimenez-Grau:2020jrx}
dimensions there is a significant body of literature, however five
\cite{Chang:2017cdx} and six \cite{Beem:2015aoa,Chang:2017xmr,Alday:2020tgi}
dimensional analyses have been carried out comparatively less. In general, the
bounds that are imposed by crossing symmetry and associativity of the OPE appear to be
saturated by SCFTs that have a construction via string theory. While most of
the literature has focused on theories in lower dimensions, the geometric
landscape of 6d $(1,0)$ SCFTs that has been charted in
\cite{Heckman:2013pva,Heckman:2015bfa} is
opportune for exploitation via bootstrap techniques.

In this paper, we use the numerical conformal bootstrap to study
six-dimensional $\mathcal{N}=(1,0)$ SCFTs with a non-Abelian flavor symmetry.
We focus on those with flavor algebra $\mathfrak{so}_{4k}$, for $k > 4$. We
find that the lower bound imposed by unitarity on the flavor central charge
for a given value of $k$ converges to that of minimal $(D_k, D_k)$
conformal matter. This suggests that such conformal matter is the ``smallest''
SCFT with that flavor symmetry, and rules out any potential more exotic theory
with a smaller flavor central charge. Assuming that the bound is saturated, we
proceed to extract the conformal dimensions of the first few scalar long
multiplets. We observe that the half-BPS multiplets that appear to be absent
from the spectrum are consistent with those known to be projected out by the
Higgs-branch chiral ring of minimal $(D_k, D_k)$ conformal matter, providing further
confirmation for the conjecture that conformal matter saturates the bounds.

\section{Minimal $(D_k, D_k)$ Conformal Matter}\label{sec:typeD_CM}

Before discussing how the conformal bootstrap can be used to learn about
the data of 6d SCFTs, we review how they can be engineered in F-theory, and in particular how minimal $(D_k,D_k)$ conformal matter and
some of its properties arise from geometry.

Six-dimensional theories with superconformal symmetry are obtained by compactification of F-theory on particular non-compact elliptically-fibered Calabi--Yau threefolds. The internal space encodes many properties of the theory, including the required cancellation of gauge anomalies.\footnote{For a recent review of the construction and properties of 6d SCFTs that arise from F-theory, see \cite{Heckman:2018jxk}.}
As mentioned in the introduction, an extra ingredient in six-dimensional
conformal theories is the presence of tensionless strings in their spectrum -- and the
tensor multiplets they magnetically couple to -- which are absent in their
lower-dimensional cousins. These can be realized by wrapping D3-branes on
curves in the base of the elliptic fibration, where the tension of the string is set by the volume of the curve. To obtain an SCFT it is therefore necessary that there are no curves of finite volume, as they would otherwise introduce a scale induced by the tension.

In practice, one can begin by enumerating all non-compact bases containing
configurations of contractible curves such that there exists a minimal elliptic
fibration over them. Compactification of F-theory on these geometries however
gives rise to 6d field theories containing tensionful strings. This geometry
corresponds to the tensor branch of a 6d SCFT if it is possible to
simultaneously shrink all rigid curves to zero volume; this can be done if the
intersection matrix of the curves -- corresponding to the Dirac pairing of the 6d
strings -- is negative definite. This condition, in addition to the requirement
that the elliptic fibration has only minimal singularities, makes it possible
to enumerate all such tensor-branch geometries leading to an SCFT
\cite{Heckman:2013pva,Heckman:2015bfa}.

In this top-down approach, minimal $(D_k, D_k)$ conformal matter is realized
through an elliptic fibration over $\mathbb{C}^2$. The elliptic fiber over a
generic point is a smooth torus, and over two divisors, $z_1$ and $z_2$, in
$\mathbb{C}^2$ there exist I$_{k-4}^{*s}$ singular fibers.\footnote{For the
notation for the types of singular fibers in elliptically-fibered Calabi--Yau
threefolds, we refer the reader to \cite{Bershadsky:1996nh}.} These singular
fibers make manifest an $\mathfrak{so}_{2k} \oplus \mathfrak{so}_{2k}$ flavor
algebra. The tensor-branch geometry is obtained by blowing up the intersection
point in $\mathbb{C}^2$ between the divisors $z_1$ and $z_2$. This procedure
introduces a compact $(-1)$-curve over which the singular fiber is
I$^{ns}_{k-4}$, corresponding to an $\mathfrak{sp}_{k-4}$ gauge
algebra.\footnote{The case $k=4$ corresponds to having no gauge algebra over
	the $(-1)$-curve; this is the geometric configuration associated to the
	E-string theory, and the flavor symmetry enhances: $\mathfrak{so}_8
\oplus \mathfrak{so}_8 \rightarrow \mathfrak{e}_8$. We will assume $k > 4$ in
this article. The superconformal bootstrap for the E-string has been studied
in \cite{Chang:2017xmr}.} This tensor-branch geometry can be compactly written
in the shorthand notation
\begin{equation}\label{eqn:TBg}
    \overset{\mathfrak{sp}_{k-4}}{1} \,.
\end{equation}
On the tensor branch the theory has the following content: one tensor multiplet; a vector multiplet in the
adjoint representation of $\mathfrak{sp}_{k-4}$; and $4k$ half-hypermultiplets
in the fundamental representation of $\mathfrak{sp}_{k-4}$. As the fundamental
representation of the gauge algebra is pseudo-real, the $4k$
half-hypermultiplets are rotated by a classical $\mathfrak{so}_{4k}$ flavor
symmetry, and this indicates that the geometrically-manifest flavor algebra is
enhanced:
\begin{equation}
    \mathfrak{so}_{2k} \oplus \mathfrak{so}_{2k} \rightarrow \mathfrak{so}_{4k} \,.
\end{equation}
This $\mathfrak{so}_{4k}$ is the flavor symmetry of the gauge theory that
exists on the generic point of the tensor branch. We can verify that
$\mathfrak{so}_{4k}$ is also the flavor symmetry of the SCFT at the origin of
the tensor branch by studying, for example, the infinite-coupling magnetic
quiver \cite{Hanany:2018uhm}.

Our knowledge of the conformal data, such as the scaling dimensions and OPE
coefficients, of the 6d SCFTs constructed from geometry in the manner that we
have just described remains limited. While the SCFT may possess a
weakly-coupled regime away from the fixed point, one of the obstacles is that
it is generally unknown how to fully track data obtained using such a
description, e.g.~the gauge-invariant operators on the tensor branch, to the
origin where the SCFT resides. From the geometry, we are however able to
compute a particular set of  symmetry-protected quantities on the tensor
branch, and follow them through the geometric deformations that lead to the
origin in a controlled way. 

One such quantity is the anomaly polynomial. Moving onto the tensor branch,
where the scalar fields inside the tensor multiplets receive non-trivial vacuum
expectation values, conformal invariance is broken.  However, since these
scalars are uncharged under the other symmetries, conformality is the only
symmetry which is broken. As such, one can use a form of 't Hooft anomaly
matching to determine the anomaly polynomial of the SCFT at the conformal fixed
point \cite{Ohmori:2014kda,Intriligator:2014eaa}. 

For a theory in $d$ dimensions, the anomaly polynomial is a formal $(d+2)$-form
which captures the variation of the partition function under the application of
a symmetry transformation via the Wess--Zumino descent procedure. In 6d, it is
generally written in terms of the characteristic classes of the bundles
associated to the gravitational, flavor, and $SU(2)_R$ symmetries; it measures
an obstruction to the gauging of these symmetries.\footnote{A recent summary on
the determination of anomaly polynomials of 6d SCFTs appears in
\cite{Baume:2021qho}. We refer to that paper for the conventions used herein.}
The $(1,0)$ supersymmetry mandates that various coefficients appearing in the
anomaly polynomial can be related to the central charges of the theory, and
these can further be related to certain OPE coefficients. The bounds derived on
the OPE coefficients from the conformal bootstrap then lead to bounds on
quantities which can be obtained geometrically from the anomaly polynomial. 

The anomaly polynomial for a 6d SCFT contains the terms
\begin{equation}\label{eqn:I8general}
	I_8 = \sum_a\text{Tr}F_a^2 \big(\kappa_a p_1(T) + \nu_a c_2(R)\big) + \cdots \,,
\end{equation}
where $p_1(T)$ is the first Pontryagin class of the tangent bundle to the
six-dimensional spacetime, $c_2(R)$ is the second Chern class of the $SU(2)_R$
R-symmetry bundle, and $\text{Tr}F_a^2$ is the curvature of each flavor
symmetry bundle. The index $a$ runs over all simple non-Abelian flavor
symmetries.

It has been shown in \cite{Chang:2017xmr,Cordova:2019wns} that the flavor
central charges, which are defined through the two-point correlation function
of the flavor currents,\footnote{There is a slight difference in normalization
with respect to \cite{Cordova:2019wns}: $C_J(\text{us}) = \text{vol}(S^5)^2\,C_J(\text{them})$.}
\begin{equation}\label{eqn:JJ}
    \left<J_\mu^a(x) J_\nu^b(0)\right> = \frac{(C_J)_a}{\text{vol}(S^5)^2} \,\delta^{ab} \frac{x^2\eta_{\mu\nu} - 2x_\mu x_\nu}{x^{12}}\,,
\end{equation}
can be written in terms of the 't Hooft coefficients of the anomaly polynomial
via
\begin{equation}\label{eqn:ccs}
    (C_J)_a = 240 \left( \kappa_a - \nu_a \right) \,.
\end{equation}
In the F-theory construction, the anomaly polynomial can be determined
from the geometric data of the associated non-compact elliptically-fibered
Calabi--Yau, together with (mixed-)gauge anomaly cancellation. For $(D_k, D_k)$ conformal matter it was worked out in
\cite{Ohmori:2014kda}, using the tensor-branch geometry as in equation
\eqref{eqn:TBg}, and the relevant coefficients from the anomaly polynomial were
found to be
\begin{equation}
      \nu = - \frac{1}{4}(k - 3) \,, \qquad 
      \kappa = \frac{1}{48}(k - 1) \,.
\end{equation}
As the theory has only a single simple flavor symmetry factor,
$\mathfrak{so}_{4k}$, we have suppressed the index $a$. From the relation
(\ref{eqn:ccs}) we can immediately see that the flavor central charge
is\footnote{The normalization differs from that of \cite{Chang:2017xmr} by a factor of two.}
\begin{equation}
    C_J = 65(k-4) + 75 \,.
\end{equation}

The flavor current associated to a flavor symmetry of a 6d $\mathcal{N}=(1,0)$
SCFT belong to the half-BPS superconformal multiplet known as a
$\mathcal{D}[2]$-multiplet \cite{Cordova:2016emh}. The highest-weight state of
this multiplet is an adjoint-valued scalar field known as the moment map,
$\phi$.\footnote{The highest-weight state of a $\mathcal{D}[2J_R]$-multiplet
transforms in the $SU(2)_R$ representation with highest weight $(2J_R)$ and has
conformal dimension $\Delta=4J_R$. Thus, the moment map transforms in the
adjoint representation of both $SU(2)_R$ and the flavor symmetry.} The OPE of
two of these moment maps contains a contribution from the $\mathcal{D}[2]$
multiplet, with OPE coefficient $\lambda_{\mathcal{D}[2]} = \lambda_{\phi\phi\mathcal{D}[2]}$. This coefficient is related to the flavor central charge \cite{Chang:2017xmr}, via the
definition in equation \eqref{eqn:JJ}:
\begin{equation}
    C_J = \frac{5h^\vee}{\lambda_{\mathcal{D}[2]}^2} \,,
\end{equation}
where $h^\vee$ is the dual Coxeter number of the flavor algebra. Upper bounds
on $\lambda_{\mathcal{D}[2]}^2$ -- and thus lower bounds on $C_J$ -- can be
determined from the superconformal bootstrap, to which we turn in Section
\ref{sec:sum_rules}.

Another interesting feature of a 6d SCFT, and one which can be gainfully
employed in a superconformal bootstrap approach, is the associated Higgs
branch. This is the hyperk\"ahler moduli space where the half-BPS states gain
vacuum expectation values and as such it is a more refined property than the
numerical value of $C_J$. It is not known how to determine the Higgs branch in
general.  If the 6d SCFT admits a Type IIA description however, then the
methods of magnetic quivers \cite{Hanany:1996ie} can be used. Fortunately,
minimal $(D_k, D_k)$ conformal matter possesses such a Type IIA description,
and their magnetic quivers were studied in \cite{Ferlito:2017xdq,
Hanany:2018uhm}.

One must distinguish the Higgs branch of the tensor-branch theory from that of
the conformal fixed point. In either case, the structure of the Higgs branch is preserved by
dimensional reduction and it can be studied using the associated 3d magnetic
quivers. In the former case, the Higgs branch is the closure of the nilpotent
orbit of $\mathfrak{so}_{4k}$ associated to the partition $[2^{2k-8},1^{16}]$
of $4k$, and the Higgs branch chiral ring is finitely-generated by the moment
map operator. Typically the chiral ring is not freely generated and the
presence of chiral-ring relations is determined from the Hilbert series,
which, in turn, can be computed from the magnetic quiver. Due to the half-BPS
states arising from the tensionless strings, the dimension of the Higgs branch
jumps by $29$ when one travels to the origin of the tensor branch. A new
generator of the chiral ring appears at the SCFT point, and it transforms in one
of the spinor representations of $\mathfrak{so}_{4k}$ with R-charge $2J_R = k-2$.

The Hilbert series (HS) encodes which of the flavor
representations are allowed to appear in chiral-ring relations involving a
spin-$J_R$ representation of $SU(2)_R$. Any flavor symmetry representation that
does not appear in the Hilbert series at order $t^{2J_R}$ implies that the
theory does not contain a $\mathcal{D}[2J_R]$ superconformal multiplet
transforming in that representation. Denoting the irreducible representations
of $\mathfrak{so}_{4k}$ appearing in $\textbf{adj}\otimes\textbf{adj}$ by
$\mathcal{R}_i$ (as in Figure \ref{fig:son}), one finds the following universal
contributions to the Hilbert series at the conformal point:
\begin{equation}\label{eqn:HS}
  \begin{aligned}
	\text{HS}(t) &= \sum_{J_R\geq0} c_{J_R} t^{2J_R} \\
	&= (\mathcal{R}_1) + (\mathcal{R}_5) t^2 + (\mathcal{R}_1 + \mathcal{R}_3 + \mathcal{R}_4 )t^4 + \cdots \,,
  \end{aligned}
\end{equation}
with $\mathcal{R}_1=\textbf{1}$ and $\mathcal{R}_5=\textbf{adj}$. The ellipses
indicate terms with $J_R>2$ or other representations that do not appear in the
$\textbf{adj}\otimes\textbf{adj}$ decomposition. For instance, the presence of
the second generator might lead to extra cubic or quadratic terms in the
Hilbert series, but these are not relevant for our purposes.

An important consequence of this analysis is that the unit operator and the
moment map can only transform in the singlet or adjoint representations
respectively, as expected, and furthermore that half-BPS states with $J_R=2$ are forbidden
to transform in the representations $\mathcal{R}_i\,,i=2,5,6$. We will see in
Section \ref{sec:result} that this selection rule will provide an additional cross-check to the claim that conformal matter saturates unitarity
bounds.

\section{The Conformal Bootstrap}\label{sec:sum_rules}

The conformal bootstrap relies on the associativity of the OPE and the
decomposition of four-point correlation functions in terms of (super)conformal
blocks to extract constraints on the spectrum. The majority of bootstrap
studies focus on correlation functions of Lorentz scalars; the structure of
their OPE and the conformal blocks being well understood in those cases
\cite{Dolan:2000ut,Dolan:2003hv, Dolan:2011dv}. As we are interested in
obtaining bounds on theories with flavor symmetry $\mathfrak{so}_{4k}$, we use that the moment map is a Lorentz scalar, as seen in Section
\ref{sec:typeD_CM}. For 6d SCFTs with eight supercharges, the sum rules used in the conformal bootstrap were first derived in \cite{Chang:2017cdx} and applied to theories with $\mathfrak{f}=\mathfrak{e}_8$ flavor symmetry, which we now review with minor modifications for the case where $\mathfrak{f}=\mathfrak{so}_{4k}$.

In order to avoid cluttering due to the proliferation of R-symmetry indices, it
is customary to introduce an auxiliary variable $Y^A\,,A=1,2$, and define
degree-two homogeneous functions, $\phi^a(x,Y)=\phi^a(x)_{AB}Y^AY^B$. The
correlation functions of four of these operators are then constrained by symmetry
to take the form \cite{Dolan:2004mu}:
\begin{equation}
\begin{aligned}
	&\left<\phi^a(x_1,Y_1)\phi^b(x_2,Y_2)\phi^c(x_3,Y_3)\phi^d(x_4,Y_4)\right> = \cr&\qquad\qquad\qquad
	\frac{(Y_1\cdot Y_2)^2(Y_3\cdot Y_4)^2}{x_{12}^8x_{34}^8} G^{abcd}(u,v;w)\,.
\end{aligned}
\end{equation}
The three variables $u,v,w$ are called the cross-ratios, and are invariant
under conformal and $SU(2)_R$ transformations:
\begin{equation}
	\begin{gathered}
	u = \frac{x_{12}^2x_{34}^2}{x_{13}^2x_{24}^2}\,,\quad
	v = \frac{x_{14}^2x_{23}^2}{x_{13}^2x_{24}^2}\,,\\
	w = \frac{(Y_1\cdot Y_2)(Y_3\cdot Y_4)}{(Y_1\cdot Y_4)(Y_2\cdot Y_3)}\,,\\
	x_{pq}^2 = |x_p-x_q|^2\,,\quad
	Y_p\cdot Y_q = \epsilon_{AB}Y^A_p Y^B_q\,.
	\end{gathered}
\end{equation}
The four-point function must also be a four-index invariant tensor of the
flavor symmetry. The conformal- and $SU(2)_R$-invariant part of
the correlation function, $G^{abcd}(u,v;w)$, can therefore further be decomposed into a
sum over the projectors onto irreducible representations
$\mathcal{R}_i$ appearing in $\textbf{adj}\otimes\textbf{adj}$ \cite{Rattazzi:2010yc}:
\begin{equation}\label{flavor-decomp}
	G^{abcd}(u,v;w) = \sum_{\mathcal{R}_i\in\textbf{adj}\otimes\textbf{adj}} P^{abcd}_i \, G_i(u,v;w)\,.
\end{equation}
The tensors $P^{abcd}_i$ are the projectors onto $\mathcal{R}_i$ and satisfy the
usual properties \cite{Cvitanovic:2008zz}: 
\begin{equation} 
	P_i^{abcd}P_j^{dcef} = \delta_{ij}P_i^{abef}\,,\qquad 
	P_i^{abba}= \text{dim}(\mathcal{R}_i)\,.
\end{equation}
Having broken down the four-point function into invariants for each of the
flavor symmetry channels $G_i(u,v;w)$, we further decompose it into
contributions from each of the superconformal multiplets, $\chi$, appearing in
the OPE of two moment maps and transforming in a given irreducible
representation $\mathcal{R}_i$ of the flavor symmetry:
\begin{equation}\label{block-decomp}
	G_i(u,v;w) = \sum_{\substack{\chi\in\phi\times\phi\\\chi \text{ in } \mathcal{R}_i}} \lambda_{\chi,\mathcal{R}_i}^2 \mathcal{G}_\chi(u,v;w)\,.
\end{equation}
For ease of notation, we write the OPE coefficients of $\chi\in\phi\times\phi$
as $\lambda_{\phi\phi\chi}=\lambda_{\chi}$. The superconformal blocks,
	$\mathcal{G}_\chi(u,v;w)$, can themselves be expanded as a linear combination over
	the non-supersymmetric conformal blocks associated to the bosonic
	primaries in the superconformal multiplet, and they satisfy both a Casimir
	differential equation \cite{Bobev:2017jhk} and a Ward identity
	\cite{Dolan:2001tt,Dolan:2004mu}. This allows one to write each
	coefficient as a rational function depending solely on the quantum
numbers of the superconformal primary. For theories with eight supercharges
and $2<d\leq6$, this analysis was performed in detail for the moment map in
\cite{Chang:2017xmr,Bobev:2017jhk} and generalized to arbitrary
$\mathcal{D}$-type half-BPS multiplets in \cite{Baume:2019aid}, to which we
refer for the exact expressions.

In addition to the form of the superconformal blocks, it was also found that
not all types of superconformal multiplets are allowed to appear in the OPE.
Let us denote a superconformal multiplet by $\chi[2J_R]_{\Delta,\ell,
\mathcal{R}}$. For 6d SCFTs with $\mathcal{N}=(1,0)$ supersymmetry, the
multiplets can be long, $\chi=\mathcal{L}$, or short,
$\chi=\mathcal{A},\mathcal{B}, \mathcal{C}, \mathcal{D}$
\cite{Buican:2016hpb,Cordova:2016emh}. In the present case, the superconformal
primary always transforms in the $\ell$-traceless symmetric representation of
the Lorentz group and has integer R-charge $J_R$.\footnote{Generically, the
superconformal multiplets depend on all three Dynkin indices of
$\mathfrak{so}_6$, but for OPEs of scalars, we are restricted to
$[0,\ell,0]$ representations.} It has conformal dimension $\Delta$,
while $\mathcal{R}$ indicates the representation under the flavor symmetry. For
short multiplets, the superconformal primary is annihilated by a particular
subset of the supercharges, fixing some of its quantum numbers. In those
cases, we drop the associated subscript. For instance,
$\mathcal{D}[2J_R]$-type superconformal primaries are half-BPS and must be
scalars ($\ell=0$) of conformal dimension $\Delta= 4J_R$. 

It turns out that $\mathcal{A}$- and $\mathcal{C}$-type multiplets cannot
appear in the decomposition in equation \eqref{block-decomp}, while long
multiplets must be R-symmetry singlets. Generically, only the following
multiplets are allowed:\footnote{Type $\mathcal{B}[0]_{\ell>0}$ multiplets are
	in principle also allowed in the OPE, but they include higher-spin
	conserved currents. The presence of these multiplets in the spectrum
	implies that (at least a subsector of) the theory is free
\cite{Maldacena:2011jn,Buican:2016hpb,Cordova:2016emh}. We exclude them as we
focus on interacting SCFTs.}
\begin{equation}\label{allowed-multiplets}
	\begin{gathered}
		\mathcal{L}[0]_{\Delta,\ell,\mathcal{R}}\,,\qquad
		\mathcal{B}[0]_{0,\mathcal{R}}\,,\qquad
		\mathcal{B}[2]_{\ell,\mathcal{R}}\,,\\
		\mathcal{D}[0]_{\bm{1}}\,,\qquad
		\mathcal{D}[2]_{\textbf{adj}}\,,\qquad
		\mathcal{D}[4]_{\mathcal{R}}\,.
	\end{gathered}
\end{equation}
The unit operator, $\mathcal{D}[0]$, and the moment-map superconformal
multiplet, $\mathcal{D}[2]$, must transform in the singlet and adjoint
representations of the flavor symmetry, respectively, as we also observed from the Hilbert series in equation \eqref{eqn:HS}. We denote a generic superconformal multiplet transforming in a representation
$\mathcal{R}$ of the flavor symmetry as $\chi_\mathcal{R}$ when
the other quantum numbers are not relevant.

Having a block decomposition of the four-point function, the important
observation that led to the conformal bootstrap is that performing an OPE in
either of the $s$, $t$, or $u$ channels does not change its structure. Using
the properties of superconformal blocks under exchange of kinematic
variables,
\begin{align}
	(1 \leftrightarrow 2): \,\, G_i(u,v;w) &= (-1)^{|\mathcal{R}_i|}\,G_i(\frac{u}{v},\frac{1}{v};\frac{-w}{w+1})\,,\label{4pt-12}\\
	(1 \leftrightarrow 3): \,\, G_i(u,v;w) &= \left( \frac{u^2}{v^2 w} \right)^{2} G_i(v,u;w^{-1})\,,\label{4pt-13}
\end{align}
one then obtains two sets of constraints from the crossing symmetry of the
four-point function \cite{Chang:2017xmr}.

In the specific case of the moment map, invariance under exchange of
$(x_1,Y_1,a)\leftrightarrow (x_2,Y_2,b)$ leads to an additional selection rule:
a superconformal multiplet can only appear in the conformal block decomposition
if its quantum numbers satisfy:
\begin{equation}\label{quantum-number-constraint}
	\ell + J_R + |\mathcal{R}_i| \in 2\mathbb{Z}\,,
\end{equation}
where $|\mathcal{R}_i|$ is defined as the parity of the embedding of
$\mathcal{R}_i$ in $\textbf{adj}\otimes\textbf{adj}$, specifically $0$ or $1$ if the representation is embedded symmetrically or anti-symmetrically, respectively. In Figure \ref{fig:son}, we give the decomposition into irreducible
representations for $\mathfrak{so}_{4k}$ and the relevant group-theoretic
quantities.

On the other hand, invariance under the exchange $(x_1,Y_1,a)\leftrightarrow
(x_3,Y_3,c)$, in combination with equations \eqref{flavor-decomp} and
\eqref{4pt-13}, leads to the following constraint:
\begin{equation}\label{sum-rules}
	F_i^{\,j} G_j(u,v;w) = \frac{u^4}{v^4w^2} G_i(v,u;w^{-1})\,.
\end{equation}
The crossing matrix, $F_i^{\,j}$, captures how the flavor representations are
reshuffled when going from the $s$ channel to the $t$ channel. The indices $i, j$ run over the irreducible representations $\mathcal{R}_i$ inside of $\textbf{adj} \otimes \textbf{adj}$. The matrix is defined via the following combination of the projectors \cite{Rattazzi:2010yc}:
\begin{equation}
	F_i^{\,j} = \frac{1}{\text{dim}(\mathcal{R}_i)} P_{i}^{dabc} P^{abcd}_j\,,\qquad F_i^{\,k}F_k^{\,j}=\delta^j_i\,.
\end{equation}
It is therefore a purely group-theoretic quantity, and for
$\mathfrak{f}=\mathfrak{so}_n$, using ``birdtrack'' techniques
\cite{Cvitanovic:2008zz}, a lengthy but straightforward computation leads to the
results collated in Figure \ref{fig:son}.

\begin{figure*}
	\centering
	\renewcommand{\arraystretch}{1.4}
	\setlength{\tabcolsep}{1pt} 
	    \begin{tabular}{ccccccccccccc}
		    $\textbf{adj}\otimes\textbf{adj}$ & = & $\mathcal{R}_1$ & $\oplus$ & $\mathcal{R}_2$ & $\oplus$ & $\mathcal{R}_3$ & $\oplus$ & $\mathcal{R}_4$ & $\oplus$ & $\mathcal{R}_5$ & $\oplus$ & $\mathcal{R}_6$\\
	        $\text{dim}\mathcal{R}_i$ & : & $1$ & + & $\frac{(n-1)(n+2)}{2}$ & + & ${\tiny\frac{(n-3)n(n+1)(n+2)}{12}}$ & + & $\frac{n(n-1)(n-2)(n-3)}{24}$ & + &$\frac{n(n-1)}{2}$ & + &$\frac{n(n+2)(n-1)(n-3)}{8}$\\
		    $|\mathcal{R}_i|$ &:&$+$&~&$+$&~&$+$&~&$+$&~&$-$&~&$-$
	\end{tabular}
	\begin{equation*}
	    F_i^{\,j} =\begin{pmatrix}
		    \frac{2}{n(n-1)} & \frac{n+2}{n} & \frac{(n-3) (n+1) (n+2)}{6 (n-1)} & \frac{(n-3) (n-2)}{12}  & 1 & \frac{(n-3) (n+2)}{4}  \\
		    \frac{2}{n(n-1)} & \frac{n^2-8}{2 (n-2) n} & \frac{(n-4) (n-3) (n+1)}{6 (n-2) (n-1)} & \frac{3-n}{6} & \frac{n-4}{2 (n-2)} & -\frac{n-3}{n-2} \\
		    \frac{2}{n(n-1)} & \frac{n-4}{(n-2) n} & \frac{n^2-6 n+11}{3 (n-2) (n-1)} & \frac{1}{6} & -\frac{1}{n-2} & -\frac{n-4}{2 (n-2)} \\
		    \frac{2}{n(n-1)} & -\frac{2 (n+2)}{(n-2) n} & \frac{(n+1) (n+2)}{3 (n-2) (n-1)} & \frac{1}{6} & \frac{2}{n-2} & -\frac{n+2}{2 (n-2)} \\
		    \frac{2}{n(n-1)} & \frac{(n-4) (n+2)}{2 (n-2) n} & -\frac{(n-3) (n+1) (n+2)}{6 (n-2) (n-1)} & \frac{n-3}{6} & \frac{1}{2} & 0 \\
		    \frac{2}{n(n-1)} & -\frac{4}{(n-2) n} & -\frac{(n-4) (n+1)}{3 (n-2) (n-1)} & -\frac{1}{6} & 0 & \frac{1}{2} \\
	    \end{pmatrix}
    \end{equation*}
	    \caption{Decomposition of $\textbf{adj}\otimes\textbf{adj}$ for $\mathfrak{so}_n$ algebras and
		the group-theoretic data relevant for the $\mathcal{N}=(1,0)$
		sum rules with flavor. The highest-weights of the representations are $\mathcal{R}_1=\bm{1}:[00\cdots]$,
		$\mathcal{R}_2:[20\cdots]$,
		$\mathcal{R}_3:[020\cdots]$,
		$\mathcal{R}_4:[00010\cdots]$,
		$\mathcal{R}_5=\textbf{adj}:[010\cdots]$, 
		and $\mathcal{R}_6:[1010\cdots]$.
		}
	\label{fig:son}
\end{figure*}

While the crossing matrix deals with the flavor symmetry, we still need to
decompose the constraint in equation \eqref{sum-rules} into each R-symmetry channel. Invariance under the 
R-symmetry forces the function $G_i(u,v;w)$ to be a degree-two
polynomial in $w^{-1}$ \cite{Dolan:2004mu}:
\begin{equation}\label{4pt-wexp}
	G_i(u,v;w) = \sum_{k=0}^2 G_i^{(k)}(u,v) w^{-k} \,.
\end{equation}
Using the relation in equation \eqref{4pt-13} one find constraints for each
power of $w$, but, as pointed out in \cite{Chang:2017xmr}, they are not
independent. Using the superconformal Ward identity, it is then possible to
find a single independent constraint for each flavor-representation channel.
These constraints are referred to as the \emph{sum rules}, and are given by:
\begin{equation}\label{sum-rule-D1}
	\sum_{\substack{\chi\in\phi\times\phi\\\chi \text{ in } \mathcal{R}_j}}
	\lambda_{\chi,\mathcal{R}_j}^2 \left(
		F^{\,j}_i\mathcal{K}_{\chi}(u,v) - \delta^j_i \mathcal{K}_{\chi}(v,u)
	\right) = 0\,,
\end{equation}
where the sum is over each multiplet that transforms in the representation
$\mathcal{R}_j\in\mathbf{adj}\otimes\mathbf{adj}$, subject to the selection
rule in equation \eqref{quantum-number-constraint}. Furthermore, we have defined the
function
\begin{equation}\label{definition-cal-K}
	\mathcal{K}_\chi(u,v) = v^4\mathcal{G}^{(2)}_\chi(u,v) - u^4\mathcal{G}^{(0)}_\chi(v,u) \,,
\end{equation}
and used a polynomial expansion of the superconformal blocks similar to that of
equation \eqref{4pt-wexp},\footnote{In the notation using the auxiliary
	R-symmetry variable, a possible decomposition for the superconformal
	blocks of a multiplet, $\chi$, is in terms of Legendre polynomials,
	$P_n$:
	$$\mathcal{G}_\chi(u,v;w)=\sum_{(\Delta,\ell,J)\in\chi}c_{\Delta,\ell,J}P_{2J}(1+2/w)g_{\Delta,\ell}\,,$$
	where $g_{\Delta,\ell}$ are the bosonic conformal blocks, and the sum
	is taken over the bosonic (conformal but not necessarily superconformal) primaries
in $\chi$. While convenient to derive the explicit expression of $\mathcal{G}_\chi$, we stress
that this basis is different from the one used in equation \eqref{4pt-wexp}.} 
\begin{equation}
	\mathcal{G}_\chi(u,v;w)=\sum_{k=0}^2\mathcal{G}^{(k)}_\chi(u,v)w^{-k}\,.
\end{equation}
We refer to \cite{Chang:2017xmr,Bobev:2017jhk, Baume:2019aid} for additional
details on the derivations of the sum rules and the form of the conformal
blocks.

\section{Bootstrapping Conformal Matter}\label{sec:result}

To extract constraints on the spectrum of the SCFT from the sum rules in equation \eqref{sum-rule-D1}, we
use the now-standard linear-functional method introduced in
\cite{Rattazzi:2008pe}, which we briefly summarize here. The interested reader
will find additional details in the reviews
\cite{Qualls:2015qjb, Rychkov:2016iqz,Simmons-Duffin:2016gjk,Poland:2018epd,
Chester:2019wfx}.

Consider the space of functions of the conformal cross-ratios, $f(u,v)$.
We may then define a functional, $\alpha_i$, for each of the flavor channels.
The space of such linear functionals can be parameterized by linear
combinations of derivatives of the function evaluated at, for instance, the
crossing-symmetric point, $u=v$:
\begin{equation}\label{definition-functional}
	\alpha^i[f] = \sum_{m,n} \alpha_{m,n}^i \left.\partial^m_u\partial^n_vf(u,v)\right|_{u=v}\,.
\end{equation}

Applying this functional to the sum rules and summing over all flavor channels,
we obtain the single constraint
\begin{equation}\label{sum-rule-2}
  \begin{gathered}
	\sum_{\substack{\chi_{\mathcal{R}_i}\in\phi\times\phi}} 
	\lambda_{\chi,\mathcal{R}_i}^2 \alpha[\chi_{\mathcal{R}_i}]=0\,,\cr 
	\alpha[\chi_{\mathcal{R}_j}] = F^{\,j}_i\alpha^i[\mathcal{K}_\chi(u,v)] - \delta^j_i \alpha^i[\mathcal{K}_\chi(v,u)] \,,
	\end{gathered}
\end{equation}
where we abuse the notation and use $\alpha[\chi_{\mathcal{R}_i}]$ to denote
the linear combination of the functionals $\alpha^i$ applied to the function
in equation \eqref{definition-cal-K} for a multiplet $\chi$ transforming in the
representation $\mathcal{R}_i$.

By unitarity, the OPE coefficients satisfy
$\lambda^2_{\chi,\mathcal{R}_i}\geq0$, and the numerical conformal bootstrap
involves searching for a functional such that:
\begin{equation}\label{sdp}
	\begin{aligned}
		&\alpha[\mathcal{D}[2]_{\textbf{adj}}] = 1\,;\\
		&\alpha[\chi_{\mathcal{R}_i}] \geq 0\,,\qquad\forall~\chi_{\mathcal{R}_i}\neq \mathcal{D}[0]_{\bm{1}}\,,\mathcal{D}[2]_{\textbf{adj}}\,;\\
		&\alpha[\mathcal{D}[0]_{\bm{1}}]\,\text{ maximized} \,.
	\end{aligned}
\end{equation}
Plugging back into in equation \eqref{sum-rule-2} and using
a convention in which the OPE coefficients of the identity and moment map are
normalized such that $\lambda^2_{\mathcal{D}[0],\bm{1}}=\text{dim}\,\mathfrak{f}$ and
$\lambda^2_{\mathcal{D}[2],\textbf{adj}}=\lambda^2_{\mathcal{D}[2]}$, we obtain
an upper bound on $\lambda^2_{\mathcal{D}[2]}$ and by extension a lower bound
on the flavor central charge:
\begin{equation}\label{OPE-bound}
	\begin{aligned}
		\lambda^2_{\mathcal{D}[2]} &\leq - \alpha[\mathcal{D}[0]_{\bm{1}}]\,\text{dim}\,\mathfrak{f}\,,\\
		C_J &\geq\frac{5h^\vee}{- \alpha[\mathcal{D}[0]_{\bm{1}}]\,\text{dim}\,\mathfrak{f}}\,,
	\end{aligned}
\end{equation}
with $h^\vee=(n-2)\,,\text{dim}\,\mathfrak{f}=\frac{1}{2}n(n-1)$ for
$\mathfrak{f}=\mathfrak{so}_n$. Similar bounds can be obtained for any OPE
coefficient by demanding the functional be normalized with respect to the relevant multiplet.

The system defined by equation \eqref{sdp} is called a semi-definite program,
and solving it is a well-defined optimization problem. In practice, we restrict
ourselves to a finite number of derivatives, up to a cutoff $2m+n\leq\Lambda$,
which captures only a portion of the space of functionals. As a functional
satisfying equation \eqref{sdp} for a given $\Lambda$ is included in the space
of functionals with $\Lambda+1$, we can only find improved results as the
cutoff is increased, and sending $\Lambda\to\infty$ will correspond to
strongest bound. 

There are nowadays standard tools to solve semi-definite programs, in
particular a numerical solver, $\texttt{SDPB}$, was specifically created for
applications to the numerical bootstrap
\cite{Simmons-Duffin:2015qma,Landry:2019qug}. In Appendix
\ref{app:implementation}, we explain how we implemented and solved the
semi-definite program numerically, leading to the results found in the next
sections.\footnote{The reader interested in raw data should feel free contact us.}

We stress that we obtain rigorous bounds: the conformal bootstrap only relies
on numerical algorithms to find the optimal coefficients $\alpha^i_{m,n}$ of
the functionals defined in equation \eqref{definition-functional} satisfying
equation \eqref{sdp}. While some standard approximations are necessary, such as
truncating the spin of the operators appearing in the OPE, we verified that, up
to the number of significant digits presented in the next sections, our results
are stable against the increase of these parameters.

\subsection{Bounds on Central Charges}

Having reviewed the sum rules for the moment map and the associated
semi-definite program defined in equation \eqref{sdp}, we have now gathered all
the necessary tools to find bounds on the flavor central charge of 6d
$\mathcal{N}=(1,0)$ SCFTs with $\mathfrak{f}=\mathfrak{so}_{4k}$.

From the geometric point of view, the theory with the smallest
$\mathfrak{so}_{4k}$ flavor symmetry has $k=5$ and corresponds to minimal $(D_5, D_5)$
conformal matter. Solving the semi-definite program in equation \eqref{sdp}, we
obtain the results shown in Figure \ref{fig:CJ-so20}. 

\begin{figure}[t]
     \centering
	\resizebox{0.49\textwidth}{!}{\input{figures/CJ-so20.pgf}}
     \caption{Bootstrap lower bounds on $C_J$ for theories with flavor symmetry
     $\mathfrak{so}_{20}$. The red lines correspond to quadratic interpolations for points with 
     either $\Lambda\geq33$ or $\Lambda\geq 35$. The derivative cutoffs are
     $\Lambda=7, 9,\dots,49,51$.}
     \label{fig:CJ-so20}
\end{figure}
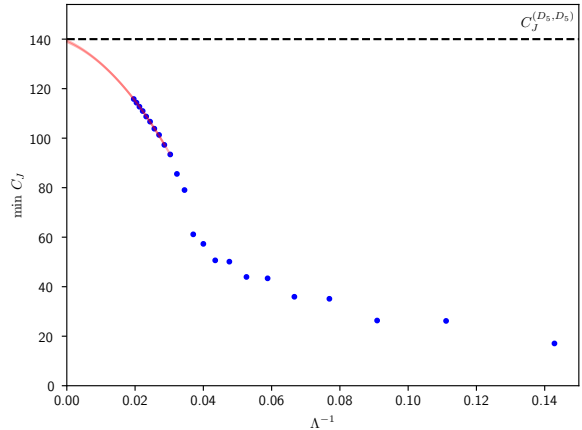
\begin{figure}[H]
     \centering
	\resizebox{0.49\textwidth}{!}{\input{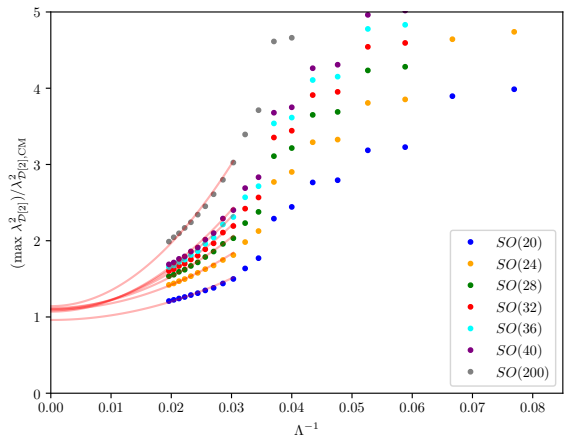}}
     \caption{Ratios between the bootstrap lower bounds on
     $\lambda^2_{\mathcal{D}[2]}$ with $\mathfrak{so}_{4k}$ flavor and the
     associated value for $(D_k,D_k)$ conformal matter. The red lines correspond to quadratic interpolations for points with $\Lambda\geq33$.}
     \label{fig:ratios}
\end{figure}

\setlength{\tabcolsep}{10pt}
\begin{table}[H]
	\centering
 	\begin{tabular}{c|c|c}
		$k$ & $\min{C_J}$\quad & $C_J^{(D_k, D_k)}$ \\\hline
 		$5$ & $115$ & $140$\\
 		$6$ & $144$ & $205$\\
 		$7$ & $175$ & $270$\\
 		$8$ & $208$ & $335$\\
 		$9$ & $241$ & $400$\\
 		$10$ & $275$ & $465$\\
 		$50$ & $1540$ & $3065$
 	\end{tabular}
	\caption{Bounds on $C_J$ for various $\mathfrak{so}_{4k}$ flavor
	symmetries, and that of $(D_k,D_k)$ conformal matter. The values for
	$\min{C_J}$ are those obtained when $\Lambda=51$.}
 	\label{tab:bounds}
 \end{table}

We can see that as the
derivative cutoff $\Lambda$ increases, the lower bound on $C_J$ improves; in
particular, with $\Lambda=51$ we obtain the strict bound $C_J>115$, ruling out
any putative spectrum with a lower value of the flavor central charge. 

Using quadratic fits on the last few points, we can further see that there is
strong evidence that as $\Lambda\to\infty$, we will obtain the bound $\min{C_J} \geq
C_J^{(D_5,D_5)} = 140$, indicating that minimal $(D_5, D_5)$ conformal matter
is the theory saturating the unitarity bounds.

We find that analogous results also hold for higher values of $k$. To aid
comparison, Figure \ref{fig:ratios} shows the ratio between the bound on
$\lambda^2_{\mathcal{D}[2]}$ and its value for minimal $(D_k,D_k)$ conformal
matter. As more derivatives are taken into account, this ratio approaches one,
again appearing to rule out any SCFTs with a smaller flavor central charge than
conformal matter.  A quadratic fit further predicts that as
$\Lambda\to\infty$, we approach $\min{C_J} = C_J^{(D_k, D_k)}$ within ten
percent.  The interpolation improves as larger values of $\Lambda$ are taken
into account, and we conjecture that minimal $(D_k,D_k)$ conformal matter saturates
the unitarity bounds for $\mathfrak{so}_{4k}$ flavor symmetry.

We emphasize again that while conformal matter has the lowest value of the
flavor central charge from the geometric point of view, it is \emph{a priori}
not obvious that, purely from superconformal-field-theoretic arguments, there
cannot exist another theory with $\mathfrak{f}=\mathfrak{so}_{4k}$, lying
outwith the F-theory construction, satisfying $C_J < C_J^{(D_k,D_k)}$. Our results
exclude a large part of those potential spectra. For instance, with $k=50$, we
rule out the existence of any theory with $C_J \leq \frac{1}{2}C_J^{(D_{50},
D_{50})}$, a bound that is even more stringent for lower values of $k$. We have
collated these bounds for $\Lambda=51$ in Table \ref{tab:bounds}.

\subsection{Low-lying Spectrum of Scalar Long Multiplets}

In addition to bounding the OPE coefficient $\lambda^2_{\mathcal{D}[2]}$, the
conformal bootstrap can also be used to extract the conformal dimension of long
multiplets appearing in the superconformal block decomposition. This is
referred to as the extremal-functional method \cite{ElShowk:2012hu}, and relies
on the fact that when the bound given in equation \eqref{OPE-bound} is
saturated, the sum rules require the associated, extremized, functional, $\alpha_E$, to
satisfy
\begin{equation}
	\alpha_E[\chi]=0\,,\qquad\forall\,\chi\neq \mathcal{D}[0]_{\bm{1}}\,,\, \mathcal{D}[2]_{\textbf{adj}}\,.
\end{equation}
Solving the constraint $\alpha_E[\mathcal{L}[0]_{\Delta,\ell,\mathcal{R}}]=0$
for a given long multiplet, we can estimate the values of conformal dimensions.
As we are operating under the assumption that the limit $\Lambda\to\infty$  corresponds to the extremal functional associated to conformal matter, for which evidence was adduced in Section \ref{sec:sum_rules}, this enables us to learn more about its spectrum.

Figures \ref{fig:L00R1}--\ref{fig:L00R4} show the functional applied to
long multiplets in various flavor representations of
$\mathfrak{f}=\mathfrak{so}_{4k}$ with $k=5, 10, 50$.
For the representations $\mathcal{R}_1,\mathcal{R}_2,\mathcal{R}_3$, there
is a gap of at least one between the dimension of lowest-lying operator and the
unitarity bound, $\Delta>6$, for long scalar multiplets.\footnote{For $k=5$,
	the vanishing of $\alpha_E[\mathcal{L}[0,0]_{\bm{1}}]$ close to
$\Delta=6$ appears to vanish for higher values of $\Lambda$.} While there are
variations, the position of the conformal dimensions does not appear to deviate
significantly as $k$ increases. 

On the other hand for $\mathcal{R}_4$, the four-antisymmetric representation, the
extremal-functional method indicates that there is an operator lying close to
threshold.  In that case, the functionals are close together around that point,
and get closer to $\Delta=6$ as $k$ grows.

At threshold, long multiplets decompose (among others) into type-$\mathcal{A}$
multiplets \cite{Buican:2016hpb,Cordova:2016emh}. As we have reviewed in
Section \ref{sec:sum_rules}, these kinds of operators are forbidden to appear
in the OPE, and the anomalous dimension therefore cannot vanish. If a multiplet
with such a small anomalous dimension is not an artifact of non-extremality,
and there is indeed a small deviation from $\Delta=6$, it would seem to
indicate the presence of a large-$k$ regime where perturbation theory can in
principle be used. This is somewhat reminiscent of large R-charge limits, which
have recently been shown to exhibit an integrable subsector
\cite{Baume:2020ure}. It would be interesting to study whether such a large-$k$
limit can be probed from the geometry, and whether there is a connection with
integrability.


\subsection{Chiral-ring Relations}

As reviewed in Section \ref{sec:typeD_CM}, $(D_k,D_k)$ conformal matter
chiral-ring relations forbid some of the $\mathcal{D}$-type superconformal
multiplets to appear in certain flavor representations. Even without solving
the semi-definite program in equation \eqref{sdp}, the selection rule in equation
\eqref{quantum-number-constraint} imposed by crossing symmetry is already
powerful enough to prevent the presence of $\mathcal{D}[4]$ multiplets in the anti-symmetric representations,
$\mathcal{R}_5,\mathcal{R}_6$, as required by the chiral ring relations, see equation \eqref{eqn:HS}.

For symmetric representations, we expect, when we approach the unitarity
bound, to find that $\alpha[\mathcal{D}[4]_\mathcal{R}]\to0$ as $\Lambda\to\infty$, if the
representation is allowed. In Table \ref{tab:chiral-ring}, we show the
numerical values of the functional for these multiplets in each of the four
symmetric representations. We can see that for $\mathcal{R}_1=\bm{1}$ and
$\mathcal{R}_3:[020\cdots]$, the functional is several orders of magnitude smaller than the
other two representations. For comparison, in the case of the multiplet
containing the stress-energy tensor, which we know appears in the OPE, we obtain
values of the same order of magnitude, $\alpha[\mathcal{B}[0,0]_{\bm{1}}]\sim
10^{-12}$. This leads us to conclude that the $\mathcal{D}[4]$ multiplet in the two-symmetric representation of the $\mathfrak{so}_{4k}$ flavor,  $\mathcal{D}[4]_{\mathcal{R}_2}$, being of order one, is forbidden to appear.

\begin{figure}[H]
     \centering
     \begin{subfigure}[b]{0.49\textwidth}
         \centering
	 \resizebox{1\textwidth}{!}{\input{figures/spectrum-so20-R1.pgf}}
	 \caption{$\mathfrak{so}_{20}$}
     \end{subfigure}
     \hfill
     \begin{subfigure}[b]{0.49\textwidth}
         \centering
	 \resizebox{1\textwidth}{!}{\input{figures/spectrum-so40-R1.pgf}}
	 \caption{$\mathfrak{so}_{40}$}
     \end{subfigure}
     \begin{subfigure}[b]{0.49\textwidth}
         \centering
	 \resizebox{1\textwidth}{!}{\input{figures/spectrum-so200-R1.pgf}}
	 \caption{$\mathfrak{so}_{200}$}
     \end{subfigure}
     \caption{$\alpha[\mathcal{L}[0]_{\Delta,0,\mathbf{1}}]$ for
     \label{fig:L00R1}
     $\Lambda=29,31,\cdots,45, 47$, from gold to purple. Zeroes of the functional indicate the presence of a long multiplet at the associated $\Delta$.}
\end{figure}
\begin{figure}[H]
     \centering
     \begin{subfigure}[b]{0.49\textwidth}
         \centering
	 \resizebox{1\textwidth}{!}{\input{figures/spectrum-so20-R2.pgf}}
	 \caption{$\mathfrak{so}_{20}$}
     \end{subfigure}
     \hfill
     \begin{subfigure}[b]{0.49\textwidth}
         \centering
	 \resizebox{1\textwidth}{!}{\input{figures/spectrum-so40-R2.pgf}}
	 \caption{$\mathfrak{so}_{40}$}
     \end{subfigure}
     \begin{subfigure}[b]{0.49\textwidth}
         \centering
	 \resizebox{1\textwidth}{!}{\input{figures/spectrum-so200-R2.pgf}}
	 \caption{$\mathfrak{so}_{200}$}
     \end{subfigure}
     \caption{$\alpha[\mathcal{L}[0]_{\Delta,0,\mathcal{R}_2}]$ for
     $\Lambda=29,31,\cdots,45, 47$, from gold to purple. Zeroes of the functional indicate the presence of a long multiplet at the associated $\Delta$.}
     \label{fig:L00R2}
\end{figure}
\begin{figure}[H]
     \centering
     \begin{subfigure}[b]{0.49\textwidth}
         \centering
	 \resizebox{1\textwidth}{!}{\input{figures/spectrum-so20-R3.pgf}}
	 \caption{$\mathfrak{so}_{20}$}
     \end{subfigure}
     \hfill
     \begin{subfigure}[b]{0.49\textwidth}
         \centering
	 \resizebox{1\textwidth}{!}{\input{figures/spectrum-so40-R3.pgf}}
	 \caption{$\mathfrak{so}_{40}$}
     \end{subfigure}
     \begin{subfigure}[b]{0.49\textwidth}
         \centering
	 \resizebox{1\textwidth}{!}{\input{figures/spectrum-so200-R3.pgf}}
	 \caption{$\mathfrak{so}_{200}$}
     \end{subfigure}
     \caption{$\alpha[\mathcal{L}[0]_{\Delta,0,\mathcal{R}_3}]$ for
     $\Lambda=29,31,\cdots,45, 47$, from gold to purple. Zeroes of the functional indicate the presence of a long multiplet at the associated $\Delta$.}
     \label{fig:L00R3}
\end{figure}
\begin{figure}[H]
     \centering
     \begin{subfigure}[b]{0.49\textwidth}
         \centering
	 \resizebox{1\textwidth}{!}{\input{figures/spectrum-so20-R4.pgf}}
	 \caption{$\mathfrak{so}_{20}$}
     \end{subfigure}
     \hfill
     \begin{subfigure}[b]{0.49\textwidth}
         \centering
	 \resizebox{1\textwidth}{!}{\input{figures/spectrum-so40-R4.pgf}}
	 \caption{$\mathfrak{so}_{40}$}
     \end{subfigure}
     \begin{subfigure}[b]{0.49\textwidth}
         \centering
	 \resizebox{1\textwidth}{!}{\input{figures/spectrum-so200-R4.pgf}}
	 \caption{$\mathfrak{so}_{200}$}
     \end{subfigure}
     \caption{$\alpha[\mathcal{L}[0]_{\Delta,0,\mathcal{R}_4}]$ for
     $\Lambda=29,31,\cdots,45, 47$, from gold to purple. Zeroes of the functional indicate the presence of a long multiplet at the associated $\Delta$.}
     \label{fig:L00R4}
\end{figure}

As we are able to reproduce the
chiral-ring condition, that $\mathcal{D}[4]$ multiplets should not appear in the representations $\mathcal{R}_2$, $\mathcal{R}_5$, or $\mathcal{R}_6$, of minimal $(D_k,D_k)$ conformal matter, this gives even more
credence to our claim that the unitarity bounds are saturated by conformal matter.

The case of $\mathcal{R}_4$, the four-antisymmetric representation, is more subtle, as the functionals seem to depend
on the value of $k$. It appears that that for low values of $k$ this
representation is forbidden but allowed for higher values.
It is very intriguing that in the case of long multiplets, the anomalous dimension
was also suppressed by $k$ for that representation. It would be interesting to
further study whether $\mathcal{R}_4$ plays an important r\^ole for conformal
matter, a question that is, to our knowledge, unexplored.

\begin{figure}[t]
     \centering
	\resizebox{0.49\textwidth}{!}{\input{figures/quarter-BPS.pgf}}
	\caption{Value of the functional with $\Lambda=49$ and
	$\mathfrak{f}=\mathfrak{so}_{200}$ applied to
	$\mathcal{B}[2,\ell]_\mathcal{R}$ multiplets. Representations not satisfying
	the constraint in equation \eqref{quantum-number-constraint} do not appear in the OPE from the outset. 
	}
     \label{fig:quarter-BPS}
\end{figure}
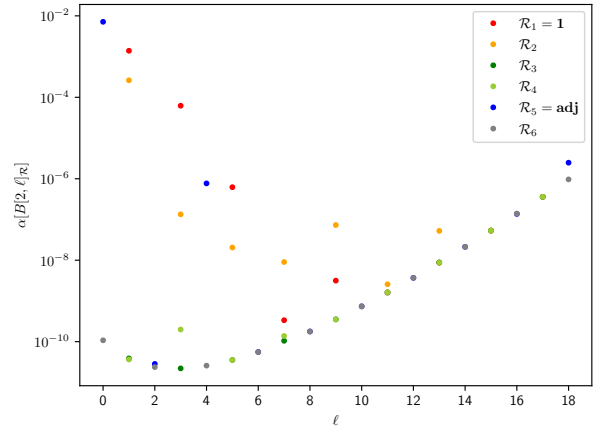

Emboldened by these results predicting the absence of half-BPS multiplets, we can endeavor to go beyond the chiral ring
and use the conformal bootstrap to predict whether there are additional
constraints related to $\mathcal{B}[2,\ell]$ operators, which must \emph{a
priori} only follow the selection rule in equation \eqref{quantum-number-constraint} and
can therefore appear in various Lorentz and flavor representations. Figure
\ref{fig:quarter-BPS} shows the value of the functional for
$\mathcal{B}[2,\ell]$ as a function of the Lorentz representation, $\ell$, for
$\mathfrak{f}=\mathfrak{so}_{200}$.

\begin{table}
	\centering
	\renewcommand{\arraystretch}{1.25}
 	\begin{tabular}{c|c|c|c|c}
		$\mathfrak{f}$	       &  $\mathcal{R}_1=\bm{1}$ &  $\mathcal{R}_2$  &  $\mathcal{R}_3$      &  $\mathcal{R}_4$\\\hline
		$\mathfrak{so}_{20}$   &  $2.7\cdot 10^{-12}$    &  $1.6$            &  $2.9\cdot 10^{-12}$  &  $2.0\cdot 10^{-1}$\\
		$\mathfrak{so}_{24}$   &  $2.0\cdot 10^{-12}$    &  $1.6$            &  $2.5\cdot 10^{-12}$  &  $4.8\cdot 10^{-2}$\\
		$\mathfrak{so}_{28}$   &  $3.6\cdot 10^{-12}$    &  $1.5$            &  $4.6\cdot 10^{-12}$  &  $2.0\cdot 10^{-2}$\\
		$\mathfrak{so}_{32}$   &  $6.3\cdot 10^{-12}$    &  $1.5$            &  $8.3\cdot 10^{-12}$  &  $1.0\cdot 10^{-2}$\\
		$\mathfrak{so}_{36}$   &  $8.5\cdot 10^{-12}$    &  $1.4$            &  $1.1\cdot 10^{-11}$  &  $6.0\cdot 10^{-3}$\\
		$\mathfrak{so}_{40}$   &  $8.2\cdot 10^{-12}$    &  $1.4$            &  $1.1\cdot 10^{-11}$  &  $3.8\cdot 10^{-3}$\\
		$\mathfrak{so}_{200}$  &  $1.6\cdot 10^{-10}$    &  $1.4$            &  $2.3\cdot 10^{-10}$  &  $1.7\cdot 10^{-5}$
 	\end{tabular}
	\caption{Values of $\alpha[\mathcal{D}[4]_\mathcal{R}]$ for $\Lambda=49$.}
 	\label{tab:chiral-ring}
 \end{table}

The value of the functional grows rapidly with $\ell$ and it becomes difficult
to comment on the presence or absence of the multiplets past the first few
values. However, the value of the functional for $\mathcal{B}[2,0]$ in the
adjoint representation is significantly larger than that of $\mathcal{R}_6$.
Similarly, $\mathcal{R}_1,\mathcal{R}_2$, the singlet and two-symmetric representations, are also orders of magnitude above the
other symmetric representations when $\ell=1,3$. It is therefore tempting
to conjecture that these multiplets are excluded from the OPE of two moment
maps. We have checked this behaviour in several cases, and there is no
indication that this potential selection rule depends on the value of $k$, and
thus it may be valid for any minimal $(D_k, D_k)$ conformal matter.  Assuming
that conformal matter saturates the bounds, it would be interesting to study
whether this conjecture on the vanishing of these particular OPE coefficients can be proven directly using either field-theoretic or
geometric techniques.

\section{Conclusions}\label{sec:conclusions}

We have explored applications of the superconformal bootstrap to 6d SCFTs with
eight supercharges, focusing on the four-point function of moment maps
associated to a flavor symmetry algebra.

We extracted bounds on the flavor central charge of theories with an
$\mathfrak{so}_{4k}$ symmetry, leaving little room for exotic theories with
smaller central charges than minimal $(D_k, D_k)$ conformal matter. In
particular, for all the explicit values of $k$ considered herein, we have
managed to exclude the existence of a consistent theory with a flavor central
charge smaller than half of that of conformal matter, a result that
significantly improves with smaller values of $k$. For instance, when $k=5$,
the smallest possible value for $(D_k, D_k)$ conformal matter, there is only
about a ten-percent window for such theories to exist. Moreover, 
quadratic interpolations reasonably show that conformal matter will saturate
the bounds imposed by unitarity as the whole space of functionals is explored.
Using the extremal-functional method, we have checked that the value of
the functional applied to half-BPS operators reproduce the expected chiral-ring
relations.

We have therefore found substantial evidence that minimal $(D_k,D_k)$ conformal matter
saturates the bounds imposed by unitarity and crossing symmetry. Thus, there cannot exist any interacting 6d SCFT, with $\mathfrak{so}_{4k}$ flavor symmetry, that has a lower value of the flavor central charge; geometry determines the extremal theory!

Assuming this is indeed correct, we have extracted the low-lying spectrum of long
operators, and also found new hints pointing to previously unknown selection
rules for conformal matter involving $\mathcal{B}$-type multiplets. For
instance, $\mathcal{B}[2,0]$ should not appear in the OPE if it transforms in
the adjoint representation, and $\mathcal{B}[2,\ell]_{\mathcal{R}_{1,2}}$
should also be forbidden when $\ell=1,3$.

Our analysis exploited a peculiarity of minimal $(D_k, D_k)$ conformal matter
in that the naive flavor symmetry,
$\mathfrak{so}_{2k}\oplus\mathfrak{so}_{2k}$, enhances to $\mathfrak{so}_{4k}$,
so that there is only one $\mathcal{D}[2]$ superconformal multiplet. Most other
types of $(\mathfrak{g},\mathfrak{g}')$ conformal matter do not possess such an
enhancement, and have two irreducible flavor currents. A natural extension of
our work is therefore to consider mixed-correlator constraints involving
multiple moment maps. Such bootstrap analyses have shown to be extremely
powerful, and often give rise to ``islands'' in parameter space.  Considering
our results, it is natural to expect those islands to correspond to minimal
$(\mathfrak{g},\mathfrak{g})$ conformal matter and their higher-rank
generalizations. Furthermore, the sum rules for mixed $\mathcal{D}[2J_R]$
correlators have been found in \cite{Baume:2019aid}. When $J_R>1$ there is more
than one independent sum rule, which should lead to more constraining results.
While the moment map is forced to transform in the adjoint representation of
the flavor symmetry, there are no such requirements for other half-BPS states.
This opens the way for flavor-independent analyses, as well as bootstrap
approaches to the whole chiral ring.

Our results have potential consequences beyond six dimensions.
Compactifying 6d $(1,0)$ SCFTs on a $T^2$ gives rise to 4d SCFTs with eight
supercharges. Starting with conformal matter one can obtain a variety
of 6d SCFTs by performing deformations and renormalization-group flows
\cite{Heckman:2015ola,Heckman:2016ssk}. In \cite{Baume:2021qho}, these Higgs-branch deformations, their compactifications on $T^2$, and their duality with
the class-$\mathcal{S}$ construction were studied; this results in a collection
of 4d $\mathcal{N}=2$ SCFTs with diverse flavor symmetry algebras for which the
central charges were determined explicitly. Compactifying minimal $(D_k, D_k)$ conformal matter on a circle, together with holonomies along the $S^1$, also leads to a vast collection of 5d SCFTs, for which the flavor symmetries were worked out in 
\cite{Apruzzi:2019vpe,Apruzzi:2019opn,Apruzzi:2019enx}. As with 6d SCFTs, not
much of the conformal data is known for these theories, although
it is reasonable to expect a dependence on the 6d progenitors, and again a conformal bootstrap approach may be useful.

Finally, we have found that the anomalous dimension of long multiplets
transforming in the four-antisymmetric representation $\mathcal{R}_4$ of $\mathfrak{so}_{4k}$
appears to be suppressed by $k$. It has recently been shown that in a large
R-charge limit, the anomalous dimensions in a particular subsector are controlled by an integrable spin chain \cite{Baume:2020ure,Heckman:2020otd}. It would be interesting to see if we can learn more about these long multiplets using perturbation theory, and whether there is an equivalent sector in the large-$k$ limit where integrability techniques can be used.

\begin{acknowledgements}

We thank Michael Fuchs for collaboration at an early stage of this work. We are
also grateful to Connor Behan, Chi-Ming Chang, Marc Gillioz, Ying-Hsuan Lin,
David Simmons-Duffin, and Alessandro Vichi for helpful discussions and insights
on the numerical implementation of the semi-definite programs. We thank Jonathan Heckman, Monica Jinwoo Kang, and Jaewon Song for comments and discussion on an early version of the manuscript. The numerical
computations in this work were performed on the Hydra cluster of the Instituto
de F\'isica Te\'orica at the Universidad Aut\'onoma de Madrid, and the General
Purpose Cluster (GPC) supported by the School of Arts and Sciences at the
University of Pennsylvania. F.~B.~is supported by the Swiss National Science
Foundation (SNSF) grant number P400P2\_194341. The work of C.~L.~was supported
by a University Research Foundation grant at the University of Pennsylvania and
DOE (HEP) Award DE-SC0021484, and by DESY (Hamburg, Germany), a member of the Helmholtz Association HGF.

\end{acknowledgements}

\appendix
\section{Numerical Implementation}\label{app:implementation}

As we have reviewed in Section \ref{sec:sum_rules}, the basic elements of the
numerical conformal bootstrap are the bosonic conformal blocks,
$g_{\Delta,\ell}(u,v)$, and their derivatives at the point $u=v$. To
solve the semi-definite program in equation \eqref{sdp}, we made use of the rational
approximation of the blocks, $\left.\partial_u^m\partial_v^n
	g_{\Delta,\ell}(u,v)\right|_{u=v}\sim
	\chi_\ell(\Delta)P^{m,n}_\ell(\Delta)$ \cite{Kos:2013tga,Penedones:2015aga}. The prefactor $\chi(\Delta)$ is
	positive for any value of the conformal dimension above the unitarity
	bound, and $P(\Delta)$ is a polynomial in the conformal dimension. The
	derivatives satisfy recursion relations found in \cite{ElShowk:2012ht,
	Kos:2013tga, Kos:2014bka} which can be efficiently utilized to find the
	rational approximation at any derivative order. We note that these relations are simpler in
	terms of the standard pair of variables $(a,b)$, see e.g.~\cite{Poland:2018epd}. In practice, we have
	therefore rewritten the sum rules in equation \eqref{sum-rule-D1} in terms of these
	variables rather than the usual cross-ratios, $(u, v)$.

Moreover, when evaluated at the crossing-symmetric point some of the derivatives of
$\mathcal{K}$ are related by a sign:
$\left.\partial^m_u\partial^n_v\mathcal{K}(u,v)\right|_{u=v} =(-1)^{m+n+1}
\left.\partial^m_u\partial^n_v\mathcal{K}(v,u)\right|_{u=v}$. This 
introduces flat directions which can lead to numerical instabilities. These can
be made manifest by rewriting the sum rule in terms of the eigenspaces of the
flavor matrix, i.e. the projectors $P^{\pm}=\frac{1}{2}(\mathbf{1}\mp
F)$, see for instance \cite{Rattazzi:2010yc,Beem:2014zpa}.

The rational approximation of the bosonic blocks and their recursion relations
have been implemented in \texttt{scalar\_blocks} 
\cite{scalarblocks}, with the value of the cut-off being related to the parameter
$n_\text{max}$ by $\Lambda = 2n_\text{max}-1$. We found the following parameters 
 adequate for $n_\text{max}\leq24$ in all the cases discussed in 
this work:
\begin{verbatim}
     poles=20
     order=80
     prec=1024
\end{verbatim}
For $n_\text{max}=25,26$, an increased precision of $\texttt{1280}$ and
$\texttt{keptPoleOrder=40}$ are needed to ensure stable results. One also needs
to introduce a cutoff for the spins, $\ell<\ell_\text{max}$, of the multiplets
appearing in the sum rules. We have tested various cases and found that at
$\ell_\text{max}=66$, the results are stable up to a sufficient number of
significant digits.

The bounds are then obtained from the semi-definite program in equation \eqref{sdp} using the
solver \texttt{sdpb} \cite{Simmons-Duffin:2015qma,Landry:2019qug} (version
\texttt{2.3.1}) with parameters:
\begin{verbatim}
    --dualityGapThreshold=1e-10
    --maxComplementarity=1e+80
    --initialMatrixScalePrimal=1e+20
    --initialMatrixScaleDual=1e+20
\end{verbatim}
The precision was the same as that used to create the bosonic blocks. The other
parameters were set to their default value. We refer to the original works and
the documentation of both \texttt{scalar\_blocks} and \texttt{sdbp} for
additional details on the numerics and the meaning of the parameters.

To test our implementation we have reproduced various results in the
literature, in particular those of six-dimensional $\mathcal{N}=(1,0)$ theories,
where we have replicated the bounds found in \cite{Chang:2017cdx} for the
E-string and a free hypermultiplet.

\bibliography{references}

\end{document}

%% file: figures/CJ-so20.pgf
\begingroup%
\makeatletter%
\begin{pgfpicture}%
\pgfpathrectangle{\pgfpointorigin}{\pgfqpoint{6.400000in}{4.800000in}}%
\pgfusepath{use as bounding box, clip}%
\begin{pgfscope}%
\pgfsetbuttcap%
\pgfsetmiterjoin%
\definecolor{currentfill}{rgb}{1.000000,1.000000,1.000000}%
\pgfsetfillcolor{currentfill}%
\pgfsetlinewidth{0.000000pt}%
\definecolor{currentstroke}{rgb}{1.000000,1.000000,1.000000}%
\pgfsetstrokecolor{currentstroke}%
\pgfsetdash{}{0pt}%
\pgfpathmoveto{\pgfqpoint{0.000000in}{0.000000in}}%
\pgfpathlineto{\pgfqpoint{6.400000in}{0.000000in}}%
\pgfpathlineto{\pgfqpoint{6.400000in}{4.800000in}}%
\pgfpathlineto{\pgfqpoint{0.000000in}{4.800000in}}%
\pgfpathclose%
\pgfusepath{fill}%
\end{pgfscope}%
\begin{pgfscope}%
\pgfsetbuttcap%
\pgfsetmiterjoin%
\definecolor{currentfill}{rgb}{1.000000,1.000000,1.000000}%
\pgfsetfillcolor{currentfill}%
\pgfsetlinewidth{0.000000pt}%
\definecolor{currentstroke}{rgb}{0.000000,0.000000,0.000000}%
\pgfsetstrokecolor{currentstroke}%
\pgfsetstrokeopacity{0.000000}%
\pgfsetdash{}{0pt}%
\pgfpathmoveto{\pgfqpoint{0.800000in}{0.528000in}}%
\pgfpathlineto{\pgfqpoint{5.760000in}{0.528000in}}%
\pgfpathlineto{\pgfqpoint{5.760000in}{4.224000in}}%
\pgfpathlineto{\pgfqpoint{0.800000in}{4.224000in}}%
\pgfpathclose%
\pgfusepath{fill}%
\end{pgfscope}%
\begin{pgfscope}%
\pgfsetbuttcap%
\pgfsetroundjoin%
\definecolor{currentfill}{rgb}{0.000000,0.000000,0.000000}%
\pgfsetfillcolor{currentfill}%
\pgfsetlinewidth{0.803000pt}%
\definecolor{currentstroke}{rgb}{0.000000,0.000000,0.000000}%
\pgfsetstrokecolor{currentstroke}%
\pgfsetdash{}{0pt}%
\pgfsys@defobject{currentmarker}{\pgfqpoint{0.000000in}{-0.048611in}}{\pgfqpoint{0.000000in}{0.000000in}}{%
\pgfpathmoveto{\pgfqpoint{0.000000in}{0.000000in}}%
\pgfpathlineto{\pgfqpoint{0.000000in}{-0.048611in}}%
\pgfusepath{stroke,fill}%
}%
\begin{pgfscope}%
\pgfsys@transformshift{0.800000in}{0.528000in}%
\pgfsys@useobject{currentmarker}{}%
\end{pgfscope}%
\end{pgfscope}%
\begin{pgfscope}%
\definecolor{textcolor}{rgb}{0.000000,0.000000,0.000000}%
\pgfsetstrokecolor{textcolor}%
\pgfsetfillcolor{textcolor}%
\pgftext[x=0.800000in,y=0.430778in,,top]{\color{textcolor}\sffamily\fontsize{10.000000}{12.000000}\selectfont 0.00}%
\end{pgfscope}%
\begin{pgfscope}%
\pgfsetbuttcap%
\pgfsetroundjoin%
\definecolor{currentfill}{rgb}{0.000000,0.000000,0.000000}%
\pgfsetfillcolor{currentfill}%
\pgfsetlinewidth{0.803000pt}%
\definecolor{currentstroke}{rgb}{0.000000,0.000000,0.000000}%
\pgfsetstrokecolor{currentstroke}%
\pgfsetdash{}{0pt}%
\pgfsys@defobject{currentmarker}{\pgfqpoint{0.000000in}{-0.048611in}}{\pgfqpoint{0.000000in}{0.000000in}}{%
\pgfpathmoveto{\pgfqpoint{0.000000in}{0.000000in}}%
\pgfpathlineto{\pgfqpoint{0.000000in}{-0.048611in}}%
\pgfusepath{stroke,fill}%
}%
\begin{pgfscope}%
\pgfsys@transformshift{1.461333in}{0.528000in}%
\pgfsys@useobject{currentmarker}{}%
\end{pgfscope}%
\end{pgfscope}%
\begin{pgfscope}%
\definecolor{textcolor}{rgb}{0.000000,0.000000,0.000000}%
\pgfsetstrokecolor{textcolor}%
\pgfsetfillcolor{textcolor}%
\pgftext[x=1.461333in,y=0.430778in,,top]{\color{textcolor}\sffamily\fontsize{10.000000}{12.000000}\selectfont 0.02}%
\end{pgfscope}%
\begin{pgfscope}%
\pgfsetbuttcap%
\pgfsetroundjoin%
\definecolor{currentfill}{rgb}{0.000000,0.000000,0.000000}%
\pgfsetfillcolor{currentfill}%
\pgfsetlinewidth{0.803000pt}%
\definecolor{currentstroke}{rgb}{0.000000,0.000000,0.000000}%
\pgfsetstrokecolor{currentstroke}%
\pgfsetdash{}{0pt}%
\pgfsys@defobject{currentmarker}{\pgfqpoint{0.000000in}{-0.048611in}}{\pgfqpoint{0.000000in}{0.000000in}}{%
\pgfpathmoveto{\pgfqpoint{0.000000in}{0.000000in}}%
\pgfpathlineto{\pgfqpoint{0.000000in}{-0.048611in}}%
\pgfusepath{stroke,fill}%
}%
\begin{pgfscope}%
\pgfsys@transformshift{2.122667in}{0.528000in}%
\pgfsys@useobject{currentmarker}{}%
\end{pgfscope}%
\end{pgfscope}%
\begin{pgfscope}%
\definecolor{textcolor}{rgb}{0.000000,0.000000,0.000000}%
\pgfsetstrokecolor{textcolor}%
\pgfsetfillcolor{textcolor}%
\pgftext[x=2.122667in,y=0.430778in,,top]{\color{textcolor}\sffamily\fontsize{10.000000}{12.000000}\selectfont 0.04}%
\end{pgfscope}%
\begin{pgfscope}%
\pgfsetbuttcap%
\pgfsetroundjoin%
\definecolor{currentfill}{rgb}{0.000000,0.000000,0.000000}%
\pgfsetfillcolor{currentfill}%
\pgfsetlinewidth{0.803000pt}%
\definecolor{currentstroke}{rgb}{0.000000,0.000000,0.000000}%
\pgfsetstrokecolor{currentstroke}%
\pgfsetdash{}{0pt}%
\pgfsys@defobject{currentmarker}{\pgfqpoint{0.000000in}{-0.048611in}}{\pgfqpoint{0.000000in}{0.000000in}}{%
\pgfpathmoveto{\pgfqpoint{0.000000in}{0.000000in}}%
\pgfpathlineto{\pgfqpoint{0.000000in}{-0.048611in}}%
\pgfusepath{stroke,fill}%
}%
\begin{pgfscope}%
\pgfsys@transformshift{2.784000in}{0.528000in}%
\pgfsys@useobject{currentmarker}{}%
\end{pgfscope}%
\end{pgfscope}%
\begin{pgfscope}%
\definecolor{textcolor}{rgb}{0.000000,0.000000,0.000000}%
\pgfsetstrokecolor{textcolor}%
\pgfsetfillcolor{textcolor}%
\pgftext[x=2.784000in,y=0.430778in,,top]{\color{textcolor}\sffamily\fontsize{10.000000}{12.000000}\selectfont 0.06}%
\end{pgfscope}%
\begin{pgfscope}%
\pgfsetbuttcap%
\pgfsetroundjoin%
\definecolor{currentfill}{rgb}{0.000000,0.000000,0.000000}%
\pgfsetfillcolor{currentfill}%
\pgfsetlinewidth{0.803000pt}%
\definecolor{currentstroke}{rgb}{0.000000,0.000000,0.000000}%
\pgfsetstrokecolor{currentstroke}%
\pgfsetdash{}{0pt}%
\pgfsys@defobject{currentmarker}{\pgfqpoint{0.000000in}{-0.048611in}}{\pgfqpoint{0.000000in}{0.000000in}}{%
\pgfpathmoveto{\pgfqpoint{0.000000in}{0.000000in}}%
\pgfpathlineto{\pgfqpoint{0.000000in}{-0.048611in}}%
\pgfusepath{stroke,fill}%
}%
\begin{pgfscope}%
\pgfsys@transformshift{3.445333in}{0.528000in}%
\pgfsys@useobject{currentmarker}{}%
\end{pgfscope}%
\end{pgfscope}%
\begin{pgfscope}%
\definecolor{textcolor}{rgb}{0.000000,0.000000,0.000000}%
\pgfsetstrokecolor{textcolor}%
\pgfsetfillcolor{textcolor}%
\pgftext[x=3.445333in,y=0.430778in,,top]{\color{textcolor}\sffamily\fontsize{10.000000}{12.000000}\selectfont 0.08}%
\end{pgfscope}%
\begin{pgfscope}%
\pgfsetbuttcap%
\pgfsetroundjoin%
\definecolor{currentfill}{rgb}{0.000000,0.000000,0.000000}%
\pgfsetfillcolor{currentfill}%
\pgfsetlinewidth{0.803000pt}%
\definecolor{currentstroke}{rgb}{0.000000,0.000000,0.000000}%
\pgfsetstrokecolor{currentstroke}%
\pgfsetdash{}{0pt}%
\pgfsys@defobject{currentmarker}{\pgfqpoint{0.000000in}{-0.048611in}}{\pgfqpoint{0.000000in}{0.000000in}}{%
\pgfpathmoveto{\pgfqpoint{0.000000in}{0.000000in}}%
\pgfpathlineto{\pgfqpoint{0.000000in}{-0.048611in}}%
\pgfusepath{stroke,fill}%
}%
\begin{pgfscope}%
\pgfsys@transformshift{4.106667in}{0.528000in}%
\pgfsys@useobject{currentmarker}{}%
\end{pgfscope}%
\end{pgfscope}%
\begin{pgfscope}%
\definecolor{textcolor}{rgb}{0.000000,0.000000,0.000000}%
\pgfsetstrokecolor{textcolor}%
\pgfsetfillcolor{textcolor}%
\pgftext[x=4.106667in,y=0.430778in,,top]{\color{textcolor}\sffamily\fontsize{10.000000}{12.000000}\selectfont 0.10}%
\end{pgfscope}%
\begin{pgfscope}%
\pgfsetbuttcap%
\pgfsetroundjoin%
\definecolor{currentfill}{rgb}{0.000000,0.000000,0.000000}%
\pgfsetfillcolor{currentfill}%
\pgfsetlinewidth{0.803000pt}%
\definecolor{currentstroke}{rgb}{0.000000,0.000000,0.000000}%
\pgfsetstrokecolor{currentstroke}%
\pgfsetdash{}{0pt}%
\pgfsys@defobject{currentmarker}{\pgfqpoint{0.000000in}{-0.048611in}}{\pgfqpoint{0.000000in}{0.000000in}}{%
\pgfpathmoveto{\pgfqpoint{0.000000in}{0.000000in}}%
\pgfpathlineto{\pgfqpoint{0.000000in}{-0.048611in}}%
\pgfusepath{stroke,fill}%
}%
\begin{pgfscope}%
\pgfsys@transformshift{4.768000in}{0.528000in}%
\pgfsys@useobject{currentmarker}{}%
\end{pgfscope}%
\end{pgfscope}%
\begin{pgfscope}%
\definecolor{textcolor}{rgb}{0.000000,0.000000,0.000000}%
\pgfsetstrokecolor{textcolor}%
\pgfsetfillcolor{textcolor}%
\pgftext[x=4.768000in,y=0.430778in,,top]{\color{textcolor}\sffamily\fontsize{10.000000}{12.000000}\selectfont 0.12}%
\end{pgfscope}%
\begin{pgfscope}%
\pgfsetbuttcap%
\pgfsetroundjoin%
\definecolor{currentfill}{rgb}{0.000000,0.000000,0.000000}%
\pgfsetfillcolor{currentfill}%
\pgfsetlinewidth{0.803000pt}%
\definecolor{currentstroke}{rgb}{0.000000,0.000000,0.000000}%
\pgfsetstrokecolor{currentstroke}%
\pgfsetdash{}{0pt}%
\pgfsys@defobject{currentmarker}{\pgfqpoint{0.000000in}{-0.048611in}}{\pgfqpoint{0.000000in}{0.000000in}}{%
\pgfpathmoveto{\pgfqpoint{0.000000in}{0.000000in}}%
\pgfpathlineto{\pgfqpoint{0.000000in}{-0.048611in}}%
\pgfusepath{stroke,fill}%
}%
\begin{pgfscope}%
\pgfsys@transformshift{5.429333in}{0.528000in}%
\pgfsys@useobject{currentmarker}{}%
\end{pgfscope}%
\end{pgfscope}%
\begin{pgfscope}%
\definecolor{textcolor}{rgb}{0.000000,0.000000,0.000000}%
\pgfsetstrokecolor{textcolor}%
\pgfsetfillcolor{textcolor}%
\pgftext[x=5.429333in,y=0.430778in,,top]{\color{textcolor}\sffamily\fontsize{10.000000}{12.000000}\selectfont 0.14}%
\end{pgfscope}%
\begin{pgfscope}%
\definecolor{textcolor}{rgb}{0.000000,0.000000,0.000000}%
\pgfsetstrokecolor{textcolor}%
\pgfsetfillcolor{textcolor}%
\pgftext[x=3.280000in,y=0.240809in,,top]{\color{textcolor}\sffamily\fontsize{10.000000}{12.000000}\selectfont \(\displaystyle \Lambda^{-1}\)}%
\end{pgfscope}%
\begin{pgfscope}%
\pgfsetbuttcap%
\pgfsetroundjoin%
\definecolor{currentfill}{rgb}{0.000000,0.000000,0.000000}%
\pgfsetfillcolor{currentfill}%
\pgfsetlinewidth{0.803000pt}%
\definecolor{currentstroke}{rgb}{0.000000,0.000000,0.000000}%
\pgfsetstrokecolor{currentstroke}%
\pgfsetdash{}{0pt}%
\pgfsys@defobject{currentmarker}{\pgfqpoint{-0.048611in}{0.000000in}}{\pgfqpoint{-0.000000in}{0.000000in}}{%
\pgfpathmoveto{\pgfqpoint{-0.000000in}{0.000000in}}%
\pgfpathlineto{\pgfqpoint{-0.048611in}{0.000000in}}%
\pgfusepath{stroke,fill}%
}%
\begin{pgfscope}%
\pgfsys@transformshift{0.800000in}{0.528000in}%
\pgfsys@useobject{currentmarker}{}%
\end{pgfscope}%
\end{pgfscope}%
\begin{pgfscope}%
\definecolor{textcolor}{rgb}{0.000000,0.000000,0.000000}%
\pgfsetstrokecolor{textcolor}%
\pgfsetfillcolor{textcolor}%
\pgftext[x=0.614412in, y=0.475238in, left, base]{\color{textcolor}\sffamily\fontsize{10.000000}{12.000000}\selectfont 0}%
\end{pgfscope}%
\begin{pgfscope}%
\pgfsetbuttcap%
\pgfsetroundjoin%
\definecolor{currentfill}{rgb}{0.000000,0.000000,0.000000}%
\pgfsetfillcolor{currentfill}%
\pgfsetlinewidth{0.803000pt}%
\definecolor{currentstroke}{rgb}{0.000000,0.000000,0.000000}%
\pgfsetstrokecolor{currentstroke}%
\pgfsetdash{}{0pt}%
\pgfsys@defobject{currentmarker}{\pgfqpoint{-0.048611in}{0.000000in}}{\pgfqpoint{-0.000000in}{0.000000in}}{%
\pgfpathmoveto{\pgfqpoint{-0.000000in}{0.000000in}}%
\pgfpathlineto{\pgfqpoint{-0.048611in}{0.000000in}}%
\pgfusepath{stroke,fill}%
}%
\begin{pgfscope}%
\pgfsys@transformshift{0.800000in}{1.008000in}%
\pgfsys@useobject{currentmarker}{}%
\end{pgfscope}%
\end{pgfscope}%
\begin{pgfscope}%
\definecolor{textcolor}{rgb}{0.000000,0.000000,0.000000}%
\pgfsetstrokecolor{textcolor}%
\pgfsetfillcolor{textcolor}%
\pgftext[x=0.526047in, y=0.955238in, left, base]{\color{textcolor}\sffamily\fontsize{10.000000}{12.000000}\selectfont 20}%
\end{pgfscope}%
\begin{pgfscope}%
\pgfsetbuttcap%
\pgfsetroundjoin%
\definecolor{currentfill}{rgb}{0.000000,0.000000,0.000000}%
\pgfsetfillcolor{currentfill}%
\pgfsetlinewidth{0.803000pt}%
\definecolor{currentstroke}{rgb}{0.000000,0.000000,0.000000}%
\pgfsetstrokecolor{currentstroke}%
\pgfsetdash{}{0pt}%
\pgfsys@defobject{currentmarker}{\pgfqpoint{-0.048611in}{0.000000in}}{\pgfqpoint{-0.000000in}{0.000000in}}{%
\pgfpathmoveto{\pgfqpoint{-0.000000in}{0.000000in}}%
\pgfpathlineto{\pgfqpoint{-0.048611in}{0.000000in}}%
\pgfusepath{stroke,fill}%
}%
\begin{pgfscope}%
\pgfsys@transformshift{0.800000in}{1.488000in}%
\pgfsys@useobject{currentmarker}{}%
\end{pgfscope}%
\end{pgfscope}%
\begin{pgfscope}%
\definecolor{textcolor}{rgb}{0.000000,0.000000,0.000000}%
\pgfsetstrokecolor{textcolor}%
\pgfsetfillcolor{textcolor}%
\pgftext[x=0.526047in, y=1.435238in, left, base]{\color{textcolor}\sffamily\fontsize{10.000000}{12.000000}\selectfont 40}%
\end{pgfscope}%
\begin{pgfscope}%
\pgfsetbuttcap%
\pgfsetroundjoin%
\definecolor{currentfill}{rgb}{0.000000,0.000000,0.000000}%
\pgfsetfillcolor{currentfill}%
\pgfsetlinewidth{0.803000pt}%
\definecolor{currentstroke}{rgb}{0.000000,0.000000,0.000000}%
\pgfsetstrokecolor{currentstroke}%
\pgfsetdash{}{0pt}%
\pgfsys@defobject{currentmarker}{\pgfqpoint{-0.048611in}{0.000000in}}{\pgfqpoint{-0.000000in}{0.000000in}}{%
\pgfpathmoveto{\pgfqpoint{-0.000000in}{0.000000in}}%
\pgfpathlineto{\pgfqpoint{-0.048611in}{0.000000in}}%
\pgfusepath{stroke,fill}%
}%
\begin{pgfscope}%
\pgfsys@transformshift{0.800000in}{1.968000in}%
\pgfsys@useobject{currentmarker}{}%
\end{pgfscope}%
\end{pgfscope}%
\begin{pgfscope}%
\definecolor{textcolor}{rgb}{0.000000,0.000000,0.000000}%
\pgfsetstrokecolor{textcolor}%
\pgfsetfillcolor{textcolor}%
\pgftext[x=0.526047in, y=1.915238in, left, base]{\color{textcolor}\sffamily\fontsize{10.000000}{12.000000}\selectfont 60}%
\end{pgfscope}%
\begin{pgfscope}%
\pgfsetbuttcap%
\pgfsetroundjoin%
\definecolor{currentfill}{rgb}{0.000000,0.000000,0.000000}%
\pgfsetfillcolor{currentfill}%
\pgfsetlinewidth{0.803000pt}%
\definecolor{currentstroke}{rgb}{0.000000,0.000000,0.000000}%
\pgfsetstrokecolor{currentstroke}%
\pgfsetdash{}{0pt}%
\pgfsys@defobject{currentmarker}{\pgfqpoint{-0.048611in}{0.000000in}}{\pgfqpoint{-0.000000in}{0.000000in}}{%
\pgfpathmoveto{\pgfqpoint{-0.000000in}{0.000000in}}%
\pgfpathlineto{\pgfqpoint{-0.048611in}{0.000000in}}%
\pgfusepath{stroke,fill}%
}%
\begin{pgfscope}%
\pgfsys@transformshift{0.800000in}{2.448000in}%
\pgfsys@useobject{currentmarker}{}%
\end{pgfscope}%
\end{pgfscope}%
\begin{pgfscope}%
\definecolor{textcolor}{rgb}{0.000000,0.000000,0.000000}%
\pgfsetstrokecolor{textcolor}%
\pgfsetfillcolor{textcolor}%
\pgftext[x=0.526047in, y=2.395238in, left, base]{\color{textcolor}\sffamily\fontsize{10.000000}{12.000000}\selectfont 80}%
\end{pgfscope}%
\begin{pgfscope}%
\pgfsetbuttcap%
\pgfsetroundjoin%
\definecolor{currentfill}{rgb}{0.000000,0.000000,0.000000}%
\pgfsetfillcolor{currentfill}%
\pgfsetlinewidth{0.803000pt}%
\definecolor{currentstroke}{rgb}{0.000000,0.000000,0.000000}%
\pgfsetstrokecolor{currentstroke}%
\pgfsetdash{}{0pt}%
\pgfsys@defobject{currentmarker}{\pgfqpoint{-0.048611in}{0.000000in}}{\pgfqpoint{-0.000000in}{0.000000in}}{%
\pgfpathmoveto{\pgfqpoint{-0.000000in}{0.000000in}}%
\pgfpathlineto{\pgfqpoint{-0.048611in}{0.000000in}}%
\pgfusepath{stroke,fill}%
}%
\begin{pgfscope}%
\pgfsys@transformshift{0.800000in}{2.928000in}%
\pgfsys@useobject{currentmarker}{}%
\end{pgfscope}%
\end{pgfscope}%
\begin{pgfscope}%
\definecolor{textcolor}{rgb}{0.000000,0.000000,0.000000}%
\pgfsetstrokecolor{textcolor}%
\pgfsetfillcolor{textcolor}%
\pgftext[x=0.437682in, y=2.875238in, left, base]{\color{textcolor}\sffamily\fontsize{10.000000}{12.000000}\selectfont 100}%
\end{pgfscope}%
\begin{pgfscope}%
\pgfsetbuttcap%
\pgfsetroundjoin%
\definecolor{currentfill}{rgb}{0.000000,0.000000,0.000000}%
\pgfsetfillcolor{currentfill}%
\pgfsetlinewidth{0.803000pt}%
\definecolor{currentstroke}{rgb}{0.000000,0.000000,0.000000}%
\pgfsetstrokecolor{currentstroke}%
\pgfsetdash{}{0pt}%
\pgfsys@defobject{currentmarker}{\pgfqpoint{-0.048611in}{0.000000in}}{\pgfqpoint{-0.000000in}{0.000000in}}{%
\pgfpathmoveto{\pgfqpoint{-0.000000in}{0.000000in}}%
\pgfpathlineto{\pgfqpoint{-0.048611in}{0.000000in}}%
\pgfusepath{stroke,fill}%
}%
\begin{pgfscope}%
\pgfsys@transformshift{0.800000in}{3.408000in}%
\pgfsys@useobject{currentmarker}{}%
\end{pgfscope}%
\end{pgfscope}%
\begin{pgfscope}%
\definecolor{textcolor}{rgb}{0.000000,0.000000,0.000000}%
\pgfsetstrokecolor{textcolor}%
\pgfsetfillcolor{textcolor}%
\pgftext[x=0.437682in, y=3.355238in, left, base]{\color{textcolor}\sffamily\fontsize{10.000000}{12.000000}\selectfont 120}%
\end{pgfscope}%
\begin{pgfscope}%
\pgfsetbuttcap%
\pgfsetroundjoin%
\definecolor{currentfill}{rgb}{0.000000,0.000000,0.000000}%
\pgfsetfillcolor{currentfill}%
\pgfsetlinewidth{0.803000pt}%
\definecolor{currentstroke}{rgb}{0.000000,0.000000,0.000000}%
\pgfsetstrokecolor{currentstroke}%
\pgfsetdash{}{0pt}%
\pgfsys@defobject{currentmarker}{\pgfqpoint{-0.048611in}{0.000000in}}{\pgfqpoint{-0.000000in}{0.000000in}}{%
\pgfpathmoveto{\pgfqpoint{-0.000000in}{0.000000in}}%
\pgfpathlineto{\pgfqpoint{-0.048611in}{0.000000in}}%
\pgfusepath{stroke,fill}%
}%
\begin{pgfscope}%
\pgfsys@transformshift{0.800000in}{3.888000in}%
\pgfsys@useobject{currentmarker}{}%
\end{pgfscope}%
\end{pgfscope}%
\begin{pgfscope}%
\definecolor{textcolor}{rgb}{0.000000,0.000000,0.000000}%
\pgfsetstrokecolor{textcolor}%
\pgfsetfillcolor{textcolor}%
\pgftext[x=0.437682in, y=3.835238in, left, base]{\color{textcolor}\sffamily\fontsize{10.000000}{12.000000}\selectfont 140}%
\end{pgfscope}%
\begin{pgfscope}%
\definecolor{textcolor}{rgb}{0.000000,0.000000,0.000000}%
\pgfsetstrokecolor{textcolor}%
\pgfsetfillcolor{textcolor}%
\pgftext[x=0.382126in,y=2.376000in,,bottom,rotate=90.000000]{\color{textcolor}\sffamily\fontsize{10.000000}{12.000000}\selectfont \(\displaystyle \min\,C_J\)}%
\end{pgfscope}%
\begin{pgfscope}%
\pgfpathrectangle{\pgfqpoint{0.800000in}{0.528000in}}{\pgfqpoint{4.960000in}{3.696000in}}%
\pgfusepath{clip}%
\pgfsetbuttcap%
\pgfsetroundjoin%
\definecolor{currentfill}{rgb}{0.000000,0.000000,1.000000}%
\pgfsetfillcolor{currentfill}%
\pgfsetlinewidth{1.003750pt}%
\definecolor{currentstroke}{rgb}{0.000000,0.000000,1.000000}%
\pgfsetstrokecolor{currentstroke}%
\pgfsetdash{}{0pt}%
\pgfsys@defobject{currentmarker}{\pgfqpoint{-0.020833in}{-0.020833in}}{\pgfqpoint{0.020833in}{0.020833in}}{%
\pgfpathmoveto{\pgfqpoint{0.000000in}{-0.020833in}}%
\pgfpathcurveto{\pgfqpoint{0.005525in}{-0.020833in}}{\pgfqpoint{0.010825in}{-0.018638in}}{\pgfqpoint{0.014731in}{-0.014731in}}%
\pgfpathcurveto{\pgfqpoint{0.018638in}{-0.010825in}}{\pgfqpoint{0.020833in}{-0.005525in}}{\pgfqpoint{0.020833in}{0.000000in}}%
\pgfpathcurveto{\pgfqpoint{0.020833in}{0.005525in}}{\pgfqpoint{0.018638in}{0.010825in}}{\pgfqpoint{0.014731in}{0.014731in}}%
\pgfpathcurveto{\pgfqpoint{0.010825in}{0.018638in}}{\pgfqpoint{0.005525in}{0.020833in}}{\pgfqpoint{0.000000in}{0.020833in}}%
\pgfpathcurveto{\pgfqpoint{-0.005525in}{0.020833in}}{\pgfqpoint{-0.010825in}{0.018638in}}{\pgfqpoint{-0.014731in}{0.014731in}}%
\pgfpathcurveto{\pgfqpoint{-0.018638in}{0.010825in}}{\pgfqpoint{-0.020833in}{0.005525in}}{\pgfqpoint{-0.020833in}{0.000000in}}%
\pgfpathcurveto{\pgfqpoint{-0.020833in}{-0.005525in}}{\pgfqpoint{-0.018638in}{-0.010825in}}{\pgfqpoint{-0.014731in}{-0.014731in}}%
\pgfpathcurveto{\pgfqpoint{-0.010825in}{-0.018638in}}{\pgfqpoint{-0.005525in}{-0.020833in}}{\pgfqpoint{0.000000in}{-0.020833in}}%
\pgfpathclose%
\pgfusepath{stroke,fill}%
}%
\begin{pgfscope}%
\pgfsys@transformshift{5.523810in}{0.937757in}%
\pgfsys@useobject{currentmarker}{}%
\end{pgfscope}%
\begin{pgfscope}%
\pgfsys@transformshift{4.474074in}{1.156348in}%
\pgfsys@useobject{currentmarker}{}%
\end{pgfscope}%
\begin{pgfscope}%
\pgfsys@transformshift{3.806061in}{1.160037in}%
\pgfsys@useobject{currentmarker}{}%
\end{pgfscope}%
\begin{pgfscope}%
\pgfsys@transformshift{3.343590in}{1.370841in}%
\pgfsys@useobject{currentmarker}{}%
\end{pgfscope}%
\begin{pgfscope}%
\pgfsys@transformshift{3.004444in}{1.390462in}%
\pgfsys@useobject{currentmarker}{}%
\end{pgfscope}%
\begin{pgfscope}%
\pgfsys@transformshift{2.745098in}{1.569072in}%
\pgfsys@useobject{currentmarker}{}%
\end{pgfscope}%
\begin{pgfscope}%
\pgfsys@transformshift{2.540351in}{1.582644in}%
\pgfsys@useobject{currentmarker}{}%
\end{pgfscope}%
\begin{pgfscope}%
\pgfsys@transformshift{2.374603in}{1.730751in}%
\pgfsys@useobject{currentmarker}{}%
\end{pgfscope}%
\begin{pgfscope}%
\pgfsys@transformshift{2.237681in}{1.743409in}%
\pgfsys@useobject{currentmarker}{}%
\end{pgfscope}%
\begin{pgfscope}%
\pgfsys@transformshift{2.122667in}{1.903190in}%
\pgfsys@useobject{currentmarker}{}%
\end{pgfscope}%
\begin{pgfscope}%
\pgfsys@transformshift{2.024691in}{1.995536in}%
\pgfsys@useobject{currentmarker}{}%
\end{pgfscope}%
\begin{pgfscope}%
\pgfsys@transformshift{1.940230in}{2.425046in}%
\pgfsys@useobject{currentmarker}{}%
\end{pgfscope}%
\begin{pgfscope}%
\pgfsys@transformshift{1.866667in}{2.581394in}%
\pgfsys@useobject{currentmarker}{}%
\end{pgfscope}%
\begin{pgfscope}%
\pgfsys@transformshift{1.802020in}{2.770053in}%
\pgfsys@useobject{currentmarker}{}%
\end{pgfscope}%
\begin{pgfscope}%
\pgfsys@transformshift{1.744762in}{2.862746in}%
\pgfsys@useobject{currentmarker}{}%
\end{pgfscope}%
\begin{pgfscope}%
\pgfsys@transformshift{1.693694in}{2.959358in}%
\pgfsys@useobject{currentmarker}{}%
\end{pgfscope}%
\begin{pgfscope}%
\pgfsys@transformshift{1.647863in}{3.019846in}%
\pgfsys@useobject{currentmarker}{}%
\end{pgfscope}%
\begin{pgfscope}%
\pgfsys@transformshift{1.606504in}{3.088202in}%
\pgfsys@useobject{currentmarker}{}%
\end{pgfscope}%
\begin{pgfscope}%
\pgfsys@transformshift{1.568992in}{3.137706in}%
\pgfsys@useobject{currentmarker}{}%
\end{pgfscope}%
\begin{pgfscope}%
\pgfsys@transformshift{1.534815in}{3.191000in}%
\pgfsys@useobject{currentmarker}{}%
\end{pgfscope}%
\begin{pgfscope}%
\pgfsys@transformshift{1.503546in}{3.233000in}%
\pgfsys@useobject{currentmarker}{}%
\end{pgfscope}%
\begin{pgfscope}%
\pgfsys@transformshift{1.474830in}{3.272073in}%
\pgfsys@useobject{currentmarker}{}%
\end{pgfscope}%
\begin{pgfscope}%
\pgfsys@transformshift{1.448366in}{3.307446in}%
\pgfsys@useobject{currentmarker}{}%
\end{pgfscope}%
\end{pgfscope}%
\begin{pgfscope}%
\pgfpathrectangle{\pgfqpoint{0.800000in}{0.528000in}}{\pgfqpoint{4.960000in}{3.696000in}}%
\pgfusepath{clip}%
\pgfsetbuttcap%
\pgfsetroundjoin%
\pgfsetlinewidth{1.505625pt}%
\definecolor{currentstroke}{rgb}{0.000000,0.000000,0.000000}%
\pgfsetstrokecolor{currentstroke}%
\pgfsetdash{{5.550000pt}{2.400000pt}}{0.000000pt}%
\pgfpathmoveto{\pgfqpoint{0.800000in}{3.888000in}}%
\pgfpathlineto{\pgfqpoint{5.760000in}{3.888000in}}%
\pgfusepath{stroke}%
\end{pgfscope}%
\begin{pgfscope}%
\pgfpathrectangle{\pgfqpoint{0.800000in}{0.528000in}}{\pgfqpoint{4.960000in}{3.696000in}}%
\pgfusepath{clip}%
\pgfsetrectcap%
\pgfsetroundjoin%
\pgfsetlinewidth{1.505625pt}%
\definecolor{currentstroke}{rgb}{1.000000,0.000000,0.000000}%
\pgfsetstrokecolor{currentstroke}%
\pgfsetstrokeopacity{0.300000}%
\pgfsetdash{}{0pt}%
\pgfpathmoveto{\pgfqpoint{0.800000in}{3.861007in}}%
\pgfpathlineto{\pgfqpoint{0.819281in}{3.852579in}}%
\pgfpathlineto{\pgfqpoint{0.838562in}{3.843656in}}%
\pgfpathlineto{\pgfqpoint{0.857843in}{3.834239in}}%
\pgfpathlineto{\pgfqpoint{0.877123in}{3.824328in}}%
\pgfpathlineto{\pgfqpoint{0.896404in}{3.813922in}}%
\pgfpathlineto{\pgfqpoint{0.915685in}{3.803023in}}%
\pgfpathlineto{\pgfqpoint{0.934966in}{3.791629in}}%
\pgfpathlineto{\pgfqpoint{0.954247in}{3.779741in}}%
\pgfpathlineto{\pgfqpoint{0.973528in}{3.767359in}}%
\pgfpathlineto{\pgfqpoint{0.992809in}{3.754482in}}%
\pgfpathlineto{\pgfqpoint{1.012089in}{3.741112in}}%
\pgfpathlineto{\pgfqpoint{1.031370in}{3.727247in}}%
\pgfpathlineto{\pgfqpoint{1.050651in}{3.712888in}}%
\pgfpathlineto{\pgfqpoint{1.069932in}{3.698035in}}%
\pgfpathlineto{\pgfqpoint{1.089213in}{3.682688in}}%
\pgfpathlineto{\pgfqpoint{1.108494in}{3.666846in}}%
\pgfpathlineto{\pgfqpoint{1.127775in}{3.650510in}}%
\pgfpathlineto{\pgfqpoint{1.147055in}{3.633680in}}%
\pgfpathlineto{\pgfqpoint{1.166336in}{3.616356in}}%
\pgfpathlineto{\pgfqpoint{1.185617in}{3.598538in}}%
\pgfpathlineto{\pgfqpoint{1.204898in}{3.580225in}}%
\pgfpathlineto{\pgfqpoint{1.224179in}{3.561418in}}%
\pgfpathlineto{\pgfqpoint{1.243460in}{3.542117in}}%
\pgfpathlineto{\pgfqpoint{1.262741in}{3.522322in}}%
\pgfpathlineto{\pgfqpoint{1.282021in}{3.502033in}}%
\pgfpathlineto{\pgfqpoint{1.301302in}{3.481249in}}%
\pgfpathlineto{\pgfqpoint{1.320583in}{3.459971in}}%
\pgfpathlineto{\pgfqpoint{1.339864in}{3.438199in}}%
\pgfpathlineto{\pgfqpoint{1.359145in}{3.415933in}}%
\pgfpathlineto{\pgfqpoint{1.378426in}{3.393172in}}%
\pgfpathlineto{\pgfqpoint{1.397707in}{3.369918in}}%
\pgfpathlineto{\pgfqpoint{1.416987in}{3.346169in}}%
\pgfpathlineto{\pgfqpoint{1.436268in}{3.321926in}}%
\pgfpathlineto{\pgfqpoint{1.455549in}{3.297189in}}%
\pgfpathlineto{\pgfqpoint{1.474830in}{3.271957in}}%
\pgfpathlineto{\pgfqpoint{1.494111in}{3.246232in}}%
\pgfpathlineto{\pgfqpoint{1.513392in}{3.220012in}}%
\pgfpathlineto{\pgfqpoint{1.532672in}{3.193298in}}%
\pgfpathlineto{\pgfqpoint{1.551953in}{3.166090in}}%
\pgfpathlineto{\pgfqpoint{1.571234in}{3.138387in}}%
\pgfpathlineto{\pgfqpoint{1.590515in}{3.110190in}}%
\pgfpathlineto{\pgfqpoint{1.609796in}{3.081500in}}%
\pgfpathlineto{\pgfqpoint{1.629077in}{3.052315in}}%
\pgfpathlineto{\pgfqpoint{1.648358in}{3.022635in}}%
\pgfpathlineto{\pgfqpoint{1.667638in}{2.992462in}}%
\pgfpathlineto{\pgfqpoint{1.686919in}{2.961794in}}%
\pgfpathlineto{\pgfqpoint{1.706200in}{2.930632in}}%
\pgfpathlineto{\pgfqpoint{1.725481in}{2.898976in}}%
\pgfpathlineto{\pgfqpoint{1.744762in}{2.866826in}}%
\pgfusepath{stroke}%
\end{pgfscope}%
\begin{pgfscope}%
\pgfpathrectangle{\pgfqpoint{0.800000in}{0.528000in}}{\pgfqpoint{4.960000in}{3.696000in}}%
\pgfusepath{clip}%
\pgfsetrectcap%
\pgfsetroundjoin%
\pgfsetlinewidth{1.505625pt}%
\definecolor{currentstroke}{rgb}{1.000000,0.000000,0.000000}%
\pgfsetstrokecolor{currentstroke}%
\pgfsetstrokeopacity{0.300000}%
\pgfsetdash{}{0pt}%
\pgfpathmoveto{\pgfqpoint{0.800000in}{3.875995in}}%
\pgfpathlineto{\pgfqpoint{0.820449in}{3.866245in}}%
\pgfpathlineto{\pgfqpoint{0.840899in}{3.855961in}}%
\pgfpathlineto{\pgfqpoint{0.861348in}{3.845142in}}%
\pgfpathlineto{\pgfqpoint{0.881798in}{3.833789in}}%
\pgfpathlineto{\pgfqpoint{0.902247in}{3.821900in}}%
\pgfpathlineto{\pgfqpoint{0.922696in}{3.809478in}}%
\pgfpathlineto{\pgfqpoint{0.943146in}{3.796521in}}%
\pgfpathlineto{\pgfqpoint{0.963595in}{3.783029in}}%
\pgfpathlineto{\pgfqpoint{0.984045in}{3.769003in}}%
\pgfpathlineto{\pgfqpoint{1.004494in}{3.754442in}}%
\pgfpathlineto{\pgfqpoint{1.024943in}{3.739347in}}%
\pgfpathlineto{\pgfqpoint{1.045393in}{3.723717in}}%
\pgfpathlineto{\pgfqpoint{1.065842in}{3.707552in}}%
\pgfpathlineto{\pgfqpoint{1.086291in}{3.690853in}}%
\pgfpathlineto{\pgfqpoint{1.106741in}{3.673619in}}%
\pgfpathlineto{\pgfqpoint{1.127190in}{3.655851in}}%
\pgfpathlineto{\pgfqpoint{1.147640in}{3.637549in}}%
\pgfpathlineto{\pgfqpoint{1.168089in}{3.618711in}}%
\pgfpathlineto{\pgfqpoint{1.188538in}{3.599340in}}%
\pgfpathlineto{\pgfqpoint{1.208988in}{3.579433in}}%
\pgfpathlineto{\pgfqpoint{1.229437in}{3.558992in}}%
\pgfpathlineto{\pgfqpoint{1.249887in}{3.538017in}}%
\pgfpathlineto{\pgfqpoint{1.270336in}{3.516507in}}%
\pgfpathlineto{\pgfqpoint{1.290785in}{3.494462in}}%
\pgfpathlineto{\pgfqpoint{1.311235in}{3.471883in}}%
\pgfpathlineto{\pgfqpoint{1.331684in}{3.448769in}}%
\pgfpathlineto{\pgfqpoint{1.352134in}{3.425121in}}%
\pgfpathlineto{\pgfqpoint{1.372583in}{3.400938in}}%
\pgfpathlineto{\pgfqpoint{1.393032in}{3.376221in}}%
\pgfpathlineto{\pgfqpoint{1.413482in}{3.350969in}}%
\pgfpathlineto{\pgfqpoint{1.433931in}{3.325183in}}%
\pgfpathlineto{\pgfqpoint{1.454381in}{3.298862in}}%
\pgfpathlineto{\pgfqpoint{1.474830in}{3.272006in}}%
\pgfpathlineto{\pgfqpoint{1.495279in}{3.244616in}}%
\pgfpathlineto{\pgfqpoint{1.515729in}{3.216691in}}%
\pgfpathlineto{\pgfqpoint{1.536178in}{3.188232in}}%
\pgfpathlineto{\pgfqpoint{1.556627in}{3.159238in}}%
\pgfpathlineto{\pgfqpoint{1.577077in}{3.129710in}}%
\pgfpathlineto{\pgfqpoint{1.597526in}{3.099647in}}%
\pgfpathlineto{\pgfqpoint{1.617976in}{3.069049in}}%
\pgfpathlineto{\pgfqpoint{1.638425in}{3.037917in}}%
\pgfpathlineto{\pgfqpoint{1.658874in}{3.006251in}}%
\pgfpathlineto{\pgfqpoint{1.679324in}{2.974050in}}%
\pgfpathlineto{\pgfqpoint{1.699773in}{2.941314in}}%
\pgfpathlineto{\pgfqpoint{1.720223in}{2.908044in}}%
\pgfpathlineto{\pgfqpoint{1.740672in}{2.874239in}}%
\pgfpathlineto{\pgfqpoint{1.761121in}{2.839900in}}%
\pgfpathlineto{\pgfqpoint{1.781571in}{2.805026in}}%
\pgfpathlineto{\pgfqpoint{1.802020in}{2.769617in}}%
\pgfusepath{stroke}%
\end{pgfscope}%
\begin{pgfscope}%
\pgfsetrectcap%
\pgfsetmiterjoin%
\pgfsetlinewidth{0.803000pt}%
\definecolor{currentstroke}{rgb}{0.000000,0.000000,0.000000}%
\pgfsetstrokecolor{currentstroke}%
\pgfsetdash{}{0pt}%
\pgfpathmoveto{\pgfqpoint{0.800000in}{0.528000in}}%
\pgfpathlineto{\pgfqpoint{0.800000in}{4.224000in}}%
\pgfusepath{stroke}%
\end{pgfscope}%
\begin{pgfscope}%
\pgfsetrectcap%
\pgfsetmiterjoin%
\pgfsetlinewidth{0.803000pt}%
\definecolor{currentstroke}{rgb}{0.000000,0.000000,0.000000}%
\pgfsetstrokecolor{currentstroke}%
\pgfsetdash{}{0pt}%
\pgfpathmoveto{\pgfqpoint{5.760000in}{0.528000in}}%
\pgfpathlineto{\pgfqpoint{5.760000in}{4.224000in}}%
\pgfusepath{stroke}%
\end{pgfscope}%
\begin{pgfscope}%
\pgfsetrectcap%
\pgfsetmiterjoin%
\pgfsetlinewidth{0.803000pt}%
\definecolor{currentstroke}{rgb}{0.000000,0.000000,0.000000}%
\pgfsetstrokecolor{currentstroke}%
\pgfsetdash{}{0pt}%
\pgfpathmoveto{\pgfqpoint{0.800000in}{0.528000in}}%
\pgfpathlineto{\pgfqpoint{5.760000in}{0.528000in}}%
\pgfusepath{stroke}%
\end{pgfscope}%
\begin{pgfscope}%
\pgfsetrectcap%
\pgfsetmiterjoin%
\pgfsetlinewidth{0.803000pt}%
\definecolor{currentstroke}{rgb}{0.000000,0.000000,0.000000}%
\pgfsetstrokecolor{currentstroke}%
\pgfsetdash{}{0pt}%
\pgfpathmoveto{\pgfqpoint{0.800000in}{4.224000in}}%
\pgfpathlineto{\pgfqpoint{5.760000in}{4.224000in}}%
\pgfusepath{stroke}%
\end{pgfscope}%
\begin{pgfscope}%
\definecolor{textcolor}{rgb}{0.000000,0.000000,0.000000}%
\pgfsetstrokecolor{textcolor}%
\pgfsetfillcolor{textcolor}%
\pgftext[x=5.197867in,y=4.008000in,left,base]{\color{textcolor}\rmfamily\fontsize{10.000000}{12.000000}\selectfont \(\displaystyle C_J^{(D_5,D_5)}\)}%
\end{pgfscope}%
\end{pgfpicture}%
\makeatother%
\endgroup%

%% file: figures/quarter-BPS.pgf
\begingroup%
\makeatletter%
\begin{pgfpicture}%
\pgfpathrectangle{\pgfpointorigin}{\pgfqpoint{6.400000in}{4.800000in}}%
\pgfusepath{use as bounding box, clip}%
\begin{pgfscope}%
\pgfsetbuttcap%
\pgfsetmiterjoin%
\definecolor{currentfill}{rgb}{1.000000,1.000000,1.000000}%
\pgfsetfillcolor{currentfill}%
\pgfsetlinewidth{0.000000pt}%
\definecolor{currentstroke}{rgb}{1.000000,1.000000,1.000000}%
\pgfsetstrokecolor{currentstroke}%
\pgfsetdash{}{0pt}%
\pgfpathmoveto{\pgfqpoint{0.000000in}{0.000000in}}%
\pgfpathlineto{\pgfqpoint{6.400000in}{0.000000in}}%
\pgfpathlineto{\pgfqpoint{6.400000in}{4.800000in}}%
\pgfpathlineto{\pgfqpoint{0.000000in}{4.800000in}}%
\pgfpathclose%
\pgfusepath{fill}%
\end{pgfscope}%
\begin{pgfscope}%
\pgfsetbuttcap%
\pgfsetmiterjoin%
\definecolor{currentfill}{rgb}{1.000000,1.000000,1.000000}%
\pgfsetfillcolor{currentfill}%
\pgfsetlinewidth{0.000000pt}%
\definecolor{currentstroke}{rgb}{0.000000,0.000000,0.000000}%
\pgfsetstrokecolor{currentstroke}%
\pgfsetstrokeopacity{0.000000}%
\pgfsetdash{}{0pt}%
\pgfpathmoveto{\pgfqpoint{0.800000in}{0.528000in}}%
\pgfpathlineto{\pgfqpoint{5.760000in}{0.528000in}}%
\pgfpathlineto{\pgfqpoint{5.760000in}{4.224000in}}%
\pgfpathlineto{\pgfqpoint{0.800000in}{4.224000in}}%
\pgfpathclose%
\pgfusepath{fill}%
\end{pgfscope}%
\begin{pgfscope}%
\pgfsetbuttcap%
\pgfsetroundjoin%
\definecolor{currentfill}{rgb}{0.000000,0.000000,0.000000}%
\pgfsetfillcolor{currentfill}%
\pgfsetlinewidth{0.803000pt}%
\definecolor{currentstroke}{rgb}{0.000000,0.000000,0.000000}%
\pgfsetstrokecolor{currentstroke}%
\pgfsetdash{}{0pt}%
\pgfsys@defobject{currentmarker}{\pgfqpoint{0.000000in}{-0.048611in}}{\pgfqpoint{0.000000in}{0.000000in}}{%
\pgfpathmoveto{\pgfqpoint{0.000000in}{0.000000in}}%
\pgfpathlineto{\pgfqpoint{0.000000in}{-0.048611in}}%
\pgfusepath{stroke,fill}%
}%
\begin{pgfscope}%
\pgfsys@transformshift{1.025455in}{0.528000in}%
\pgfsys@useobject{currentmarker}{}%
\end{pgfscope}%
\end{pgfscope}%
\begin{pgfscope}%
\definecolor{textcolor}{rgb}{0.000000,0.000000,0.000000}%
\pgfsetstrokecolor{textcolor}%
\pgfsetfillcolor{textcolor}%
\pgftext[x=1.025455in,y=0.430778in,,top]{\color{textcolor}\sffamily\fontsize{10.000000}{12.000000}\selectfont 0}%
\end{pgfscope}%
\begin{pgfscope}%
\pgfsetbuttcap%
\pgfsetroundjoin%
\definecolor{currentfill}{rgb}{0.000000,0.000000,0.000000}%
\pgfsetfillcolor{currentfill}%
\pgfsetlinewidth{0.803000pt}%
\definecolor{currentstroke}{rgb}{0.000000,0.000000,0.000000}%
\pgfsetstrokecolor{currentstroke}%
\pgfsetdash{}{0pt}%
\pgfsys@defobject{currentmarker}{\pgfqpoint{0.000000in}{-0.048611in}}{\pgfqpoint{0.000000in}{0.000000in}}{%
\pgfpathmoveto{\pgfqpoint{0.000000in}{0.000000in}}%
\pgfpathlineto{\pgfqpoint{0.000000in}{-0.048611in}}%
\pgfusepath{stroke,fill}%
}%
\begin{pgfscope}%
\pgfsys@transformshift{1.526465in}{0.528000in}%
\pgfsys@useobject{currentmarker}{}%
\end{pgfscope}%
\end{pgfscope}%
\begin{pgfscope}%
\definecolor{textcolor}{rgb}{0.000000,0.000000,0.000000}%
\pgfsetstrokecolor{textcolor}%
\pgfsetfillcolor{textcolor}%
\pgftext[x=1.526465in,y=0.430778in,,top]{\color{textcolor}\sffamily\fontsize{10.000000}{12.000000}\selectfont 2}%
\end{pgfscope}%
\begin{pgfscope}%
\pgfsetbuttcap%
\pgfsetroundjoin%
\definecolor{currentfill}{rgb}{0.000000,0.000000,0.000000}%
\pgfsetfillcolor{currentfill}%
\pgfsetlinewidth{0.803000pt}%
\definecolor{currentstroke}{rgb}{0.000000,0.000000,0.000000}%
\pgfsetstrokecolor{currentstroke}%
\pgfsetdash{}{0pt}%
\pgfsys@defobject{currentmarker}{\pgfqpoint{0.000000in}{-0.048611in}}{\pgfqpoint{0.000000in}{0.000000in}}{%
\pgfpathmoveto{\pgfqpoint{0.000000in}{0.000000in}}%
\pgfpathlineto{\pgfqpoint{0.000000in}{-0.048611in}}%
\pgfusepath{stroke,fill}%
}%
\begin{pgfscope}%
\pgfsys@transformshift{2.027475in}{0.528000in}%
\pgfsys@useobject{currentmarker}{}%
\end{pgfscope}%
\end{pgfscope}%
\begin{pgfscope}%
\definecolor{textcolor}{rgb}{0.000000,0.000000,0.000000}%
\pgfsetstrokecolor{textcolor}%
\pgfsetfillcolor{textcolor}%
\pgftext[x=2.027475in,y=0.430778in,,top]{\color{textcolor}\sffamily\fontsize{10.000000}{12.000000}\selectfont 4}%
\end{pgfscope}%
\begin{pgfscope}%
\pgfsetbuttcap%
\pgfsetroundjoin%
\definecolor{currentfill}{rgb}{0.000000,0.000000,0.000000}%
\pgfsetfillcolor{currentfill}%
\pgfsetlinewidth{0.803000pt}%
\definecolor{currentstroke}{rgb}{0.000000,0.000000,0.000000}%
\pgfsetstrokecolor{currentstroke}%
\pgfsetdash{}{0pt}%
\pgfsys@defobject{currentmarker}{\pgfqpoint{0.000000in}{-0.048611in}}{\pgfqpoint{0.000000in}{0.000000in}}{%
\pgfpathmoveto{\pgfqpoint{0.000000in}{0.000000in}}%
\pgfpathlineto{\pgfqpoint{0.000000in}{-0.048611in}}%
\pgfusepath{stroke,fill}%
}%
\begin{pgfscope}%
\pgfsys@transformshift{2.528485in}{0.528000in}%
\pgfsys@useobject{currentmarker}{}%
\end{pgfscope}%
\end{pgfscope}%
\begin{pgfscope}%
\definecolor{textcolor}{rgb}{0.000000,0.000000,0.000000}%
\pgfsetstrokecolor{textcolor}%
\pgfsetfillcolor{textcolor}%
\pgftext[x=2.528485in,y=0.430778in,,top]{\color{textcolor}\sffamily\fontsize{10.000000}{12.000000}\selectfont 6}%
\end{pgfscope}%
\begin{pgfscope}%
\pgfsetbuttcap%
\pgfsetroundjoin%
\definecolor{currentfill}{rgb}{0.000000,0.000000,0.000000}%
\pgfsetfillcolor{currentfill}%
\pgfsetlinewidth{0.803000pt}%
\definecolor{currentstroke}{rgb}{0.000000,0.000000,0.000000}%
\pgfsetstrokecolor{currentstroke}%
\pgfsetdash{}{0pt}%
\pgfsys@defobject{currentmarker}{\pgfqpoint{0.000000in}{-0.048611in}}{\pgfqpoint{0.000000in}{0.000000in}}{%
\pgfpathmoveto{\pgfqpoint{0.000000in}{0.000000in}}%
\pgfpathlineto{\pgfqpoint{0.000000in}{-0.048611in}}%
\pgfusepath{stroke,fill}%
}%
\begin{pgfscope}%
\pgfsys@transformshift{3.029495in}{0.528000in}%
\pgfsys@useobject{currentmarker}{}%
\end{pgfscope}%
\end{pgfscope}%
\begin{pgfscope}%
\definecolor{textcolor}{rgb}{0.000000,0.000000,0.000000}%
\pgfsetstrokecolor{textcolor}%
\pgfsetfillcolor{textcolor}%
\pgftext[x=3.029495in,y=0.430778in,,top]{\color{textcolor}\sffamily\fontsize{10.000000}{12.000000}\selectfont 8}%
\end{pgfscope}%
\begin{pgfscope}%
\pgfsetbuttcap%
\pgfsetroundjoin%
\definecolor{currentfill}{rgb}{0.000000,0.000000,0.000000}%
\pgfsetfillcolor{currentfill}%
\pgfsetlinewidth{0.803000pt}%
\definecolor{currentstroke}{rgb}{0.000000,0.000000,0.000000}%
\pgfsetstrokecolor{currentstroke}%
\pgfsetdash{}{0pt}%
\pgfsys@defobject{currentmarker}{\pgfqpoint{0.000000in}{-0.048611in}}{\pgfqpoint{0.000000in}{0.000000in}}{%
\pgfpathmoveto{\pgfqpoint{0.000000in}{0.000000in}}%
\pgfpathlineto{\pgfqpoint{0.000000in}{-0.048611in}}%
\pgfusepath{stroke,fill}%
}%
\begin{pgfscope}%
\pgfsys@transformshift{3.530505in}{0.528000in}%
\pgfsys@useobject{currentmarker}{}%
\end{pgfscope}%
\end{pgfscope}%
\begin{pgfscope}%
\definecolor{textcolor}{rgb}{0.000000,0.000000,0.000000}%
\pgfsetstrokecolor{textcolor}%
\pgfsetfillcolor{textcolor}%
\pgftext[x=3.530505in,y=0.430778in,,top]{\color{textcolor}\sffamily\fontsize{10.000000}{12.000000}\selectfont 10}%
\end{pgfscope}%
\begin{pgfscope}%
\pgfsetbuttcap%
\pgfsetroundjoin%
\definecolor{currentfill}{rgb}{0.000000,0.000000,0.000000}%
\pgfsetfillcolor{currentfill}%
\pgfsetlinewidth{0.803000pt}%
\definecolor{currentstroke}{rgb}{0.000000,0.000000,0.000000}%
\pgfsetstrokecolor{currentstroke}%
\pgfsetdash{}{0pt}%
\pgfsys@defobject{currentmarker}{\pgfqpoint{0.000000in}{-0.048611in}}{\pgfqpoint{0.000000in}{0.000000in}}{%
\pgfpathmoveto{\pgfqpoint{0.000000in}{0.000000in}}%
\pgfpathlineto{\pgfqpoint{0.000000in}{-0.048611in}}%
\pgfusepath{stroke,fill}%
}%
\begin{pgfscope}%
\pgfsys@transformshift{4.031515in}{0.528000in}%
\pgfsys@useobject{currentmarker}{}%
\end{pgfscope}%
\end{pgfscope}%
\begin{pgfscope}%
\definecolor{textcolor}{rgb}{0.000000,0.000000,0.000000}%
\pgfsetstrokecolor{textcolor}%
\pgfsetfillcolor{textcolor}%
\pgftext[x=4.031515in,y=0.430778in,,top]{\color{textcolor}\sffamily\fontsize{10.000000}{12.000000}\selectfont 12}%
\end{pgfscope}%
\begin{pgfscope}%
\pgfsetbuttcap%
\pgfsetroundjoin%
\definecolor{currentfill}{rgb}{0.000000,0.000000,0.000000}%
\pgfsetfillcolor{currentfill}%
\pgfsetlinewidth{0.803000pt}%
\definecolor{currentstroke}{rgb}{0.000000,0.000000,0.000000}%
\pgfsetstrokecolor{currentstroke}%
\pgfsetdash{}{0pt}%
\pgfsys@defobject{currentmarker}{\pgfqpoint{0.000000in}{-0.048611in}}{\pgfqpoint{0.000000in}{0.000000in}}{%
\pgfpathmoveto{\pgfqpoint{0.000000in}{0.000000in}}%
\pgfpathlineto{\pgfqpoint{0.000000in}{-0.048611in}}%
\pgfusepath{stroke,fill}%
}%
\begin{pgfscope}%
\pgfsys@transformshift{4.532525in}{0.528000in}%
\pgfsys@useobject{currentmarker}{}%
\end{pgfscope}%
\end{pgfscope}%
\begin{pgfscope}%
\definecolor{textcolor}{rgb}{0.000000,0.000000,0.000000}%
\pgfsetstrokecolor{textcolor}%
\pgfsetfillcolor{textcolor}%
\pgftext[x=4.532525in,y=0.430778in,,top]{\color{textcolor}\sffamily\fontsize{10.000000}{12.000000}\selectfont 14}%
\end{pgfscope}%
\begin{pgfscope}%
\pgfsetbuttcap%
\pgfsetroundjoin%
\definecolor{currentfill}{rgb}{0.000000,0.000000,0.000000}%
\pgfsetfillcolor{currentfill}%
\pgfsetlinewidth{0.803000pt}%
\definecolor{currentstroke}{rgb}{0.000000,0.000000,0.000000}%
\pgfsetstrokecolor{currentstroke}%
\pgfsetdash{}{0pt}%
\pgfsys@defobject{currentmarker}{\pgfqpoint{0.000000in}{-0.048611in}}{\pgfqpoint{0.000000in}{0.000000in}}{%
\pgfpathmoveto{\pgfqpoint{0.000000in}{0.000000in}}%
\pgfpathlineto{\pgfqpoint{0.000000in}{-0.048611in}}%
\pgfusepath{stroke,fill}%
}%
\begin{pgfscope}%
\pgfsys@transformshift{5.033535in}{0.528000in}%
\pgfsys@useobject{currentmarker}{}%
\end{pgfscope}%
\end{pgfscope}%
\begin{pgfscope}%
\definecolor{textcolor}{rgb}{0.000000,0.000000,0.000000}%
\pgfsetstrokecolor{textcolor}%
\pgfsetfillcolor{textcolor}%
\pgftext[x=5.033535in,y=0.430778in,,top]{\color{textcolor}\sffamily\fontsize{10.000000}{12.000000}\selectfont 16}%
\end{pgfscope}%
\begin{pgfscope}%
\pgfsetbuttcap%
\pgfsetroundjoin%
\definecolor{currentfill}{rgb}{0.000000,0.000000,0.000000}%
\pgfsetfillcolor{currentfill}%
\pgfsetlinewidth{0.803000pt}%
\definecolor{currentstroke}{rgb}{0.000000,0.000000,0.000000}%
\pgfsetstrokecolor{currentstroke}%
\pgfsetdash{}{0pt}%
\pgfsys@defobject{currentmarker}{\pgfqpoint{0.000000in}{-0.048611in}}{\pgfqpoint{0.000000in}{0.000000in}}{%
\pgfpathmoveto{\pgfqpoint{0.000000in}{0.000000in}}%
\pgfpathlineto{\pgfqpoint{0.000000in}{-0.048611in}}%
\pgfusepath{stroke,fill}%
}%
\begin{pgfscope}%
\pgfsys@transformshift{5.534545in}{0.528000in}%
\pgfsys@useobject{currentmarker}{}%
\end{pgfscope}%
\end{pgfscope}%
\begin{pgfscope}%
\definecolor{textcolor}{rgb}{0.000000,0.000000,0.000000}%
\pgfsetstrokecolor{textcolor}%
\pgfsetfillcolor{textcolor}%
\pgftext[x=5.534545in,y=0.430778in,,top]{\color{textcolor}\sffamily\fontsize{10.000000}{12.000000}\selectfont 18}%
\end{pgfscope}%
\begin{pgfscope}%
\definecolor{textcolor}{rgb}{0.000000,0.000000,0.000000}%
\pgfsetstrokecolor{textcolor}%
\pgfsetfillcolor{textcolor}%
\pgftext[x=3.280000in,y=0.240809in,,top]{\color{textcolor}\sffamily\fontsize{10.000000}{12.000000}\selectfont \(\displaystyle \ell\)}%
\end{pgfscope}%
\begin{pgfscope}%
\pgfsetbuttcap%
\pgfsetroundjoin%
\definecolor{currentfill}{rgb}{0.000000,0.000000,0.000000}%
\pgfsetfillcolor{currentfill}%
\pgfsetlinewidth{0.803000pt}%
\definecolor{currentstroke}{rgb}{0.000000,0.000000,0.000000}%
\pgfsetstrokecolor{currentstroke}%
\pgfsetdash{}{0pt}%
\pgfsys@defobject{currentmarker}{\pgfqpoint{-0.048611in}{0.000000in}}{\pgfqpoint{-0.000000in}{0.000000in}}{%
\pgfpathmoveto{\pgfqpoint{-0.000000in}{0.000000in}}%
\pgfpathlineto{\pgfqpoint{-0.048611in}{0.000000in}}%
\pgfusepath{stroke,fill}%
}%
\begin{pgfscope}%
\pgfsys@transformshift{0.800000in}{0.955197in}%
\pgfsys@useobject{currentmarker}{}%
\end{pgfscope}%
\end{pgfscope}%
\begin{pgfscope}%
\definecolor{textcolor}{rgb}{0.000000,0.000000,0.000000}%
\pgfsetstrokecolor{textcolor}%
\pgfsetfillcolor{textcolor}%
\pgftext[x=0.359412in, y=0.902436in, left, base]{\color{textcolor}\sffamily\fontsize{10.000000}{12.000000}\selectfont \(\displaystyle {10^{-10}}\)}%
\end{pgfscope}%
\begin{pgfscope}%
\pgfsetbuttcap%
\pgfsetroundjoin%
\definecolor{currentfill}{rgb}{0.000000,0.000000,0.000000}%
\pgfsetfillcolor{currentfill}%
\pgfsetlinewidth{0.803000pt}%
\definecolor{currentstroke}{rgb}{0.000000,0.000000,0.000000}%
\pgfsetstrokecolor{currentstroke}%
\pgfsetdash{}{0pt}%
\pgfsys@defobject{currentmarker}{\pgfqpoint{-0.048611in}{0.000000in}}{\pgfqpoint{-0.000000in}{0.000000in}}{%
\pgfpathmoveto{\pgfqpoint{-0.000000in}{0.000000in}}%
\pgfpathlineto{\pgfqpoint{-0.048611in}{0.000000in}}%
\pgfusepath{stroke,fill}%
}%
\begin{pgfscope}%
\pgfsys@transformshift{0.800000in}{1.744836in}%
\pgfsys@useobject{currentmarker}{}%
\end{pgfscope}%
\end{pgfscope}%
\begin{pgfscope}%
\definecolor{textcolor}{rgb}{0.000000,0.000000,0.000000}%
\pgfsetstrokecolor{textcolor}%
\pgfsetfillcolor{textcolor}%
\pgftext[x=0.414775in, y=1.692074in, left, base]{\color{textcolor}\sffamily\fontsize{10.000000}{12.000000}\selectfont \(\displaystyle {10^{-8}}\)}%
\end{pgfscope}%
\begin{pgfscope}%
\pgfsetbuttcap%
\pgfsetroundjoin%
\definecolor{currentfill}{rgb}{0.000000,0.000000,0.000000}%
\pgfsetfillcolor{currentfill}%
\pgfsetlinewidth{0.803000pt}%
\definecolor{currentstroke}{rgb}{0.000000,0.000000,0.000000}%
\pgfsetstrokecolor{currentstroke}%
\pgfsetdash{}{0pt}%
\pgfsys@defobject{currentmarker}{\pgfqpoint{-0.048611in}{0.000000in}}{\pgfqpoint{-0.000000in}{0.000000in}}{%
\pgfpathmoveto{\pgfqpoint{-0.000000in}{0.000000in}}%
\pgfpathlineto{\pgfqpoint{-0.048611in}{0.000000in}}%
\pgfusepath{stroke,fill}%
}%
\begin{pgfscope}%
\pgfsys@transformshift{0.800000in}{2.534474in}%
\pgfsys@useobject{currentmarker}{}%
\end{pgfscope}%
\end{pgfscope}%
\begin{pgfscope}%
\definecolor{textcolor}{rgb}{0.000000,0.000000,0.000000}%
\pgfsetstrokecolor{textcolor}%
\pgfsetfillcolor{textcolor}%
\pgftext[x=0.414775in, y=2.481712in, left, base]{\color{textcolor}\sffamily\fontsize{10.000000}{12.000000}\selectfont \(\displaystyle {10^{-6}}\)}%
\end{pgfscope}%
\begin{pgfscope}%
\pgfsetbuttcap%
\pgfsetroundjoin%
\definecolor{currentfill}{rgb}{0.000000,0.000000,0.000000}%
\pgfsetfillcolor{currentfill}%
\pgfsetlinewidth{0.803000pt}%
\definecolor{currentstroke}{rgb}{0.000000,0.000000,0.000000}%
\pgfsetstrokecolor{currentstroke}%
\pgfsetdash{}{0pt}%
\pgfsys@defobject{currentmarker}{\pgfqpoint{-0.048611in}{0.000000in}}{\pgfqpoint{-0.000000in}{0.000000in}}{%
\pgfpathmoveto{\pgfqpoint{-0.000000in}{0.000000in}}%
\pgfpathlineto{\pgfqpoint{-0.048611in}{0.000000in}}%
\pgfusepath{stroke,fill}%
}%
\begin{pgfscope}%
\pgfsys@transformshift{0.800000in}{3.324112in}%
\pgfsys@useobject{currentmarker}{}%
\end{pgfscope}%
\end{pgfscope}%
\begin{pgfscope}%
\definecolor{textcolor}{rgb}{0.000000,0.000000,0.000000}%
\pgfsetstrokecolor{textcolor}%
\pgfsetfillcolor{textcolor}%
\pgftext[x=0.414775in, y=3.271350in, left, base]{\color{textcolor}\sffamily\fontsize{10.000000}{12.000000}\selectfont \(\displaystyle {10^{-4}}\)}%
\end{pgfscope}%
\begin{pgfscope}%
\pgfsetbuttcap%
\pgfsetroundjoin%
\definecolor{currentfill}{rgb}{0.000000,0.000000,0.000000}%
\pgfsetfillcolor{currentfill}%
\pgfsetlinewidth{0.803000pt}%
\definecolor{currentstroke}{rgb}{0.000000,0.000000,0.000000}%
\pgfsetstrokecolor{currentstroke}%
\pgfsetdash{}{0pt}%
\pgfsys@defobject{currentmarker}{\pgfqpoint{-0.048611in}{0.000000in}}{\pgfqpoint{-0.000000in}{0.000000in}}{%
\pgfpathmoveto{\pgfqpoint{-0.000000in}{0.000000in}}%
\pgfpathlineto{\pgfqpoint{-0.048611in}{0.000000in}}%
\pgfusepath{stroke,fill}%
}%
\begin{pgfscope}%
\pgfsys@transformshift{0.800000in}{4.113750in}%
\pgfsys@useobject{currentmarker}{}%
\end{pgfscope}%
\end{pgfscope}%
\begin{pgfscope}%
\definecolor{textcolor}{rgb}{0.000000,0.000000,0.000000}%
\pgfsetstrokecolor{textcolor}%
\pgfsetfillcolor{textcolor}%
\pgftext[x=0.414775in, y=4.060988in, left, base]{\color{textcolor}\sffamily\fontsize{10.000000}{12.000000}\selectfont \(\displaystyle {10^{-2}}\)}%
\end{pgfscope}%
\begin{pgfscope}%
\definecolor{textcolor}{rgb}{0.000000,0.000000,0.000000}%
\pgfsetstrokecolor{textcolor}%
\pgfsetfillcolor{textcolor}%
\pgftext[x=0.303857in,y=2.376000in,,bottom,rotate=90.000000]{\color{textcolor}\sffamily\fontsize{10.000000}{12.000000}\selectfont \(\displaystyle \alpha[B[2,\ell]_\mathcal{R}]\)}%
\end{pgfscope}%
\begin{pgfscope}%
\pgfpathrectangle{\pgfqpoint{0.800000in}{0.528000in}}{\pgfqpoint{4.960000in}{3.696000in}}%
\pgfusepath{clip}%
\pgfsetbuttcap%
\pgfsetroundjoin%
\definecolor{currentfill}{rgb}{1.000000,0.000000,0.000000}%
\pgfsetfillcolor{currentfill}%
\pgfsetlinewidth{1.003750pt}%
\definecolor{currentstroke}{rgb}{1.000000,0.000000,0.000000}%
\pgfsetstrokecolor{currentstroke}%
\pgfsetdash{}{0pt}%
\pgfsys@defobject{currentmarker}{\pgfqpoint{-0.020833in}{-0.020833in}}{\pgfqpoint{0.020833in}{0.020833in}}{%
\pgfpathmoveto{\pgfqpoint{0.000000in}{-0.020833in}}%
\pgfpathcurveto{\pgfqpoint{0.005525in}{-0.020833in}}{\pgfqpoint{0.010825in}{-0.018638in}}{\pgfqpoint{0.014731in}{-0.014731in}}%
\pgfpathcurveto{\pgfqpoint{0.018638in}{-0.010825in}}{\pgfqpoint{0.020833in}{-0.005525in}}{\pgfqpoint{0.020833in}{0.000000in}}%
\pgfpathcurveto{\pgfqpoint{0.020833in}{0.005525in}}{\pgfqpoint{0.018638in}{0.010825in}}{\pgfqpoint{0.014731in}{0.014731in}}%
\pgfpathcurveto{\pgfqpoint{0.010825in}{0.018638in}}{\pgfqpoint{0.005525in}{0.020833in}}{\pgfqpoint{0.000000in}{0.020833in}}%
\pgfpathcurveto{\pgfqpoint{-0.005525in}{0.020833in}}{\pgfqpoint{-0.010825in}{0.018638in}}{\pgfqpoint{-0.014731in}{0.014731in}}%
\pgfpathcurveto{\pgfqpoint{-0.018638in}{0.010825in}}{\pgfqpoint{-0.020833in}{0.005525in}}{\pgfqpoint{-0.020833in}{0.000000in}}%
\pgfpathcurveto{\pgfqpoint{-0.020833in}{-0.005525in}}{\pgfqpoint{-0.018638in}{-0.010825in}}{\pgfqpoint{-0.014731in}{-0.014731in}}%
\pgfpathcurveto{\pgfqpoint{-0.010825in}{-0.018638in}}{\pgfqpoint{-0.005525in}{-0.020833in}}{\pgfqpoint{0.000000in}{-0.020833in}}%
\pgfpathclose%
\pgfusepath{stroke,fill}%
}%
\begin{pgfscope}%
\pgfsys@transformshift{1.275960in}{3.775289in}%
\pgfsys@useobject{currentmarker}{}%
\end{pgfscope}%
\begin{pgfscope}%
\pgfsys@transformshift{1.776970in}{3.242480in}%
\pgfsys@useobject{currentmarker}{}%
\end{pgfscope}%
\begin{pgfscope}%
\pgfsys@transformshift{2.277980in}{2.452786in}%
\pgfsys@useobject{currentmarker}{}%
\end{pgfscope}%
\begin{pgfscope}%
\pgfsys@transformshift{2.778990in}{1.163183in}%
\pgfsys@useobject{currentmarker}{}%
\end{pgfscope}%
\begin{pgfscope}%
\pgfsys@transformshift{3.280000in}{1.546747in}%
\pgfsys@useobject{currentmarker}{}%
\end{pgfscope}%
\begin{pgfscope}%
\pgfsys@transformshift{3.781010in}{1.432332in}%
\pgfsys@useobject{currentmarker}{}%
\end{pgfscope}%
\begin{pgfscope}%
\pgfsys@transformshift{4.282020in}{1.721496in}%
\pgfsys@useobject{currentmarker}{}%
\end{pgfscope}%
\begin{pgfscope}%
\pgfsys@transformshift{4.783030in}{2.031892in}%
\pgfsys@useobject{currentmarker}{}%
\end{pgfscope}%
\begin{pgfscope}%
\pgfsys@transformshift{5.284040in}{2.359364in}%
\pgfsys@useobject{currentmarker}{}%
\end{pgfscope}%
\end{pgfscope}%
\begin{pgfscope}%
\pgfpathrectangle{\pgfqpoint{0.800000in}{0.528000in}}{\pgfqpoint{4.960000in}{3.696000in}}%
\pgfusepath{clip}%
\pgfsetbuttcap%
\pgfsetroundjoin%
\definecolor{currentfill}{rgb}{1.000000,0.647059,0.000000}%
\pgfsetfillcolor{currentfill}%
\pgfsetlinewidth{1.003750pt}%
\definecolor{currentstroke}{rgb}{1.000000,0.647059,0.000000}%
\pgfsetstrokecolor{currentstroke}%
\pgfsetdash{}{0pt}%
\pgfsys@defobject{currentmarker}{\pgfqpoint{-0.020833in}{-0.020833in}}{\pgfqpoint{0.020833in}{0.020833in}}{%
\pgfpathmoveto{\pgfqpoint{0.000000in}{-0.020833in}}%
\pgfpathcurveto{\pgfqpoint{0.005525in}{-0.020833in}}{\pgfqpoint{0.010825in}{-0.018638in}}{\pgfqpoint{0.014731in}{-0.014731in}}%
\pgfpathcurveto{\pgfqpoint{0.018638in}{-0.010825in}}{\pgfqpoint{0.020833in}{-0.005525in}}{\pgfqpoint{0.020833in}{0.000000in}}%
\pgfpathcurveto{\pgfqpoint{0.020833in}{0.005525in}}{\pgfqpoint{0.018638in}{0.010825in}}{\pgfqpoint{0.014731in}{0.014731in}}%
\pgfpathcurveto{\pgfqpoint{0.010825in}{0.018638in}}{\pgfqpoint{0.005525in}{0.020833in}}{\pgfqpoint{0.000000in}{0.020833in}}%
\pgfpathcurveto{\pgfqpoint{-0.005525in}{0.020833in}}{\pgfqpoint{-0.010825in}{0.018638in}}{\pgfqpoint{-0.014731in}{0.014731in}}%
\pgfpathcurveto{\pgfqpoint{-0.018638in}{0.010825in}}{\pgfqpoint{-0.020833in}{0.005525in}}{\pgfqpoint{-0.020833in}{0.000000in}}%
\pgfpathcurveto{\pgfqpoint{-0.020833in}{-0.005525in}}{\pgfqpoint{-0.018638in}{-0.010825in}}{\pgfqpoint{-0.014731in}{-0.014731in}}%
\pgfpathcurveto{\pgfqpoint{-0.010825in}{-0.018638in}}{\pgfqpoint{-0.005525in}{-0.020833in}}{\pgfqpoint{0.000000in}{-0.020833in}}%
\pgfpathclose%
\pgfusepath{stroke,fill}%
}%
\begin{pgfscope}%
\pgfsys@transformshift{1.275960in}{3.489220in}%
\pgfsys@useobject{currentmarker}{}%
\end{pgfscope}%
\begin{pgfscope}%
\pgfsys@transformshift{1.776970in}{2.189690in}%
\pgfsys@useobject{currentmarker}{}%
\end{pgfscope}%
\begin{pgfscope}%
\pgfsys@transformshift{2.277980in}{1.868106in}%
\pgfsys@useobject{currentmarker}{}%
\end{pgfscope}%
\begin{pgfscope}%
\pgfsys@transformshift{2.778990in}{1.727142in}%
\pgfsys@useobject{currentmarker}{}%
\end{pgfscope}%
\begin{pgfscope}%
\pgfsys@transformshift{3.280000in}{2.085784in}%
\pgfsys@useobject{currentmarker}{}%
\end{pgfscope}%
\begin{pgfscope}%
\pgfsys@transformshift{3.781010in}{1.511555in}%
\pgfsys@useobject{currentmarker}{}%
\end{pgfscope}%
\begin{pgfscope}%
\pgfsys@transformshift{4.282020in}{2.029146in}%
\pgfsys@useobject{currentmarker}{}%
\end{pgfscope}%
\begin{pgfscope}%
\pgfsys@transformshift{4.783030in}{2.031798in}%
\pgfsys@useobject{currentmarker}{}%
\end{pgfscope}%
\begin{pgfscope}%
\pgfsys@transformshift{5.284040in}{2.359301in}%
\pgfsys@useobject{currentmarker}{}%
\end{pgfscope}%
\end{pgfscope}%
\begin{pgfscope}%
\pgfpathrectangle{\pgfqpoint{0.800000in}{0.528000in}}{\pgfqpoint{4.960000in}{3.696000in}}%
\pgfusepath{clip}%
\pgfsetbuttcap%
\pgfsetroundjoin%
\definecolor{currentfill}{rgb}{0.000000,0.501961,0.000000}%
\pgfsetfillcolor{currentfill}%
\pgfsetlinewidth{1.003750pt}%
\definecolor{currentstroke}{rgb}{0.000000,0.501961,0.000000}%
\pgfsetstrokecolor{currentstroke}%
\pgfsetdash{}{0pt}%
\pgfsys@defobject{currentmarker}{\pgfqpoint{-0.020833in}{-0.020833in}}{\pgfqpoint{0.020833in}{0.020833in}}{%
\pgfpathmoveto{\pgfqpoint{0.000000in}{-0.020833in}}%
\pgfpathcurveto{\pgfqpoint{0.005525in}{-0.020833in}}{\pgfqpoint{0.010825in}{-0.018638in}}{\pgfqpoint{0.014731in}{-0.014731in}}%
\pgfpathcurveto{\pgfqpoint{0.018638in}{-0.010825in}}{\pgfqpoint{0.020833in}{-0.005525in}}{\pgfqpoint{0.020833in}{0.000000in}}%
\pgfpathcurveto{\pgfqpoint{0.020833in}{0.005525in}}{\pgfqpoint{0.018638in}{0.010825in}}{\pgfqpoint{0.014731in}{0.014731in}}%
\pgfpathcurveto{\pgfqpoint{0.010825in}{0.018638in}}{\pgfqpoint{0.005525in}{0.020833in}}{\pgfqpoint{0.000000in}{0.020833in}}%
\pgfpathcurveto{\pgfqpoint{-0.005525in}{0.020833in}}{\pgfqpoint{-0.010825in}{0.018638in}}{\pgfqpoint{-0.014731in}{0.014731in}}%
\pgfpathcurveto{\pgfqpoint{-0.018638in}{0.010825in}}{\pgfqpoint{-0.020833in}{0.005525in}}{\pgfqpoint{-0.020833in}{0.000000in}}%
\pgfpathcurveto{\pgfqpoint{-0.020833in}{-0.005525in}}{\pgfqpoint{-0.018638in}{-0.010825in}}{\pgfqpoint{-0.014731in}{-0.014731in}}%
\pgfpathcurveto{\pgfqpoint{-0.010825in}{-0.018638in}}{\pgfqpoint{-0.005525in}{-0.020833in}}{\pgfqpoint{0.000000in}{-0.020833in}}%
\pgfpathclose%
\pgfusepath{stroke,fill}%
}%
\begin{pgfscope}%
\pgfsys@transformshift{1.275960in}{0.792781in}%
\pgfsys@useobject{currentmarker}{}%
\end{pgfscope}%
\begin{pgfscope}%
\pgfsys@transformshift{1.776970in}{0.696000in}%
\pgfsys@useobject{currentmarker}{}%
\end{pgfscope}%
\begin{pgfscope}%
\pgfsys@transformshift{2.277980in}{0.777669in}%
\pgfsys@useobject{currentmarker}{}%
\end{pgfscope}%
\begin{pgfscope}%
\pgfsys@transformshift{2.778990in}{0.963398in}%
\pgfsys@useobject{currentmarker}{}%
\end{pgfscope}%
\begin{pgfscope}%
\pgfsys@transformshift{3.280000in}{1.171331in}%
\pgfsys@useobject{currentmarker}{}%
\end{pgfscope}%
\begin{pgfscope}%
\pgfsys@transformshift{3.781010in}{1.431809in}%
\pgfsys@useobject{currentmarker}{}%
\end{pgfscope}%
\begin{pgfscope}%
\pgfsys@transformshift{4.282020in}{1.722909in}%
\pgfsys@useobject{currentmarker}{}%
\end{pgfscope}%
\begin{pgfscope}%
\pgfsys@transformshift{4.783030in}{2.031716in}%
\pgfsys@useobject{currentmarker}{}%
\end{pgfscope}%
\begin{pgfscope}%
\pgfsys@transformshift{5.284040in}{2.359231in}%
\pgfsys@useobject{currentmarker}{}%
\end{pgfscope}%
\end{pgfscope}%
\begin{pgfscope}%
\pgfpathrectangle{\pgfqpoint{0.800000in}{0.528000in}}{\pgfqpoint{4.960000in}{3.696000in}}%
\pgfusepath{clip}%
\pgfsetbuttcap%
\pgfsetroundjoin%
\definecolor{currentfill}{rgb}{0.603922,0.803922,0.196078}%
\pgfsetfillcolor{currentfill}%
\pgfsetlinewidth{1.003750pt}%
\definecolor{currentstroke}{rgb}{0.603922,0.803922,0.196078}%
\pgfsetstrokecolor{currentstroke}%
\pgfsetdash{}{0pt}%
\pgfsys@defobject{currentmarker}{\pgfqpoint{-0.020833in}{-0.020833in}}{\pgfqpoint{0.020833in}{0.020833in}}{%
\pgfpathmoveto{\pgfqpoint{0.000000in}{-0.020833in}}%
\pgfpathcurveto{\pgfqpoint{0.005525in}{-0.020833in}}{\pgfqpoint{0.010825in}{-0.018638in}}{\pgfqpoint{0.014731in}{-0.014731in}}%
\pgfpathcurveto{\pgfqpoint{0.018638in}{-0.010825in}}{\pgfqpoint{0.020833in}{-0.005525in}}{\pgfqpoint{0.020833in}{0.000000in}}%
\pgfpathcurveto{\pgfqpoint{0.020833in}{0.005525in}}{\pgfqpoint{0.018638in}{0.010825in}}{\pgfqpoint{0.014731in}{0.014731in}}%
\pgfpathcurveto{\pgfqpoint{0.010825in}{0.018638in}}{\pgfqpoint{0.005525in}{0.020833in}}{\pgfqpoint{0.000000in}{0.020833in}}%
\pgfpathcurveto{\pgfqpoint{-0.005525in}{0.020833in}}{\pgfqpoint{-0.010825in}{0.018638in}}{\pgfqpoint{-0.014731in}{0.014731in}}%
\pgfpathcurveto{\pgfqpoint{-0.018638in}{0.010825in}}{\pgfqpoint{-0.020833in}{0.005525in}}{\pgfqpoint{-0.020833in}{0.000000in}}%
\pgfpathcurveto{\pgfqpoint{-0.020833in}{-0.005525in}}{\pgfqpoint{-0.018638in}{-0.010825in}}{\pgfqpoint{-0.014731in}{-0.014731in}}%
\pgfpathcurveto{\pgfqpoint{-0.010825in}{-0.018638in}}{\pgfqpoint{-0.005525in}{-0.020833in}}{\pgfqpoint{0.000000in}{-0.020833in}}%
\pgfpathclose%
\pgfusepath{stroke,fill}%
}%
\begin{pgfscope}%
\pgfsys@transformshift{1.275960in}{0.782537in}%
\pgfsys@useobject{currentmarker}{}%
\end{pgfscope}%
\begin{pgfscope}%
\pgfsys@transformshift{1.776970in}{1.073168in}%
\pgfsys@useobject{currentmarker}{}%
\end{pgfscope}%
\begin{pgfscope}%
\pgfsys@transformshift{2.277980in}{0.777716in}%
\pgfsys@useobject{currentmarker}{}%
\end{pgfscope}%
\begin{pgfscope}%
\pgfsys@transformshift{2.778990in}{1.010386in}%
\pgfsys@useobject{currentmarker}{}%
\end{pgfscope}%
\begin{pgfscope}%
\pgfsys@transformshift{3.280000in}{1.170319in}%
\pgfsys@useobject{currentmarker}{}%
\end{pgfscope}%
\begin{pgfscope}%
\pgfsys@transformshift{3.781010in}{1.433296in}%
\pgfsys@useobject{currentmarker}{}%
\end{pgfscope}%
\begin{pgfscope}%
\pgfsys@transformshift{4.282020in}{1.721104in}%
\pgfsys@useobject{currentmarker}{}%
\end{pgfscope}%
\begin{pgfscope}%
\pgfsys@transformshift{4.783030in}{2.031718in}%
\pgfsys@useobject{currentmarker}{}%
\end{pgfscope}%
\begin{pgfscope}%
\pgfsys@transformshift{5.284040in}{2.359232in}%
\pgfsys@useobject{currentmarker}{}%
\end{pgfscope}%
\end{pgfscope}%
\begin{pgfscope}%
\pgfpathrectangle{\pgfqpoint{0.800000in}{0.528000in}}{\pgfqpoint{4.960000in}{3.696000in}}%
\pgfusepath{clip}%
\pgfsetbuttcap%
\pgfsetroundjoin%
\definecolor{currentfill}{rgb}{0.000000,0.000000,1.000000}%
\pgfsetfillcolor{currentfill}%
\pgfsetlinewidth{1.003750pt}%
\definecolor{currentstroke}{rgb}{0.000000,0.000000,1.000000}%
\pgfsetstrokecolor{currentstroke}%
\pgfsetdash{}{0pt}%
\pgfsys@defobject{currentmarker}{\pgfqpoint{-0.020833in}{-0.020833in}}{\pgfqpoint{0.020833in}{0.020833in}}{%
\pgfpathmoveto{\pgfqpoint{0.000000in}{-0.020833in}}%
\pgfpathcurveto{\pgfqpoint{0.005525in}{-0.020833in}}{\pgfqpoint{0.010825in}{-0.018638in}}{\pgfqpoint{0.014731in}{-0.014731in}}%
\pgfpathcurveto{\pgfqpoint{0.018638in}{-0.010825in}}{\pgfqpoint{0.020833in}{-0.005525in}}{\pgfqpoint{0.020833in}{0.000000in}}%
\pgfpathcurveto{\pgfqpoint{0.020833in}{0.005525in}}{\pgfqpoint{0.018638in}{0.010825in}}{\pgfqpoint{0.014731in}{0.014731in}}%
\pgfpathcurveto{\pgfqpoint{0.010825in}{0.018638in}}{\pgfqpoint{0.005525in}{0.020833in}}{\pgfqpoint{0.000000in}{0.020833in}}%
\pgfpathcurveto{\pgfqpoint{-0.005525in}{0.020833in}}{\pgfqpoint{-0.010825in}{0.018638in}}{\pgfqpoint{-0.014731in}{0.014731in}}%
\pgfpathcurveto{\pgfqpoint{-0.018638in}{0.010825in}}{\pgfqpoint{-0.020833in}{0.005525in}}{\pgfqpoint{-0.020833in}{0.000000in}}%
\pgfpathcurveto{\pgfqpoint{-0.020833in}{-0.005525in}}{\pgfqpoint{-0.018638in}{-0.010825in}}{\pgfqpoint{-0.014731in}{-0.014731in}}%
\pgfpathcurveto{\pgfqpoint{-0.010825in}{-0.018638in}}{\pgfqpoint{-0.005525in}{-0.020833in}}{\pgfqpoint{0.000000in}{-0.020833in}}%
\pgfpathclose%
\pgfusepath{stroke,fill}%
}%
\begin{pgfscope}%
\pgfsys@transformshift{1.025455in}{4.056000in}%
\pgfsys@useobject{currentmarker}{}%
\end{pgfscope}%
\begin{pgfscope}%
\pgfsys@transformshift{1.526465in}{0.739201in}%
\pgfsys@useobject{currentmarker}{}%
\end{pgfscope}%
\begin{pgfscope}%
\pgfsys@transformshift{2.027475in}{2.489612in}%
\pgfsys@useobject{currentmarker}{}%
\end{pgfscope}%
\begin{pgfscope}%
\pgfsys@transformshift{2.528485in}{0.855024in}%
\pgfsys@useobject{currentmarker}{}%
\end{pgfscope}%
\begin{pgfscope}%
\pgfsys@transformshift{3.029495in}{1.054266in}%
\pgfsys@useobject{currentmarker}{}%
\end{pgfscope}%
\begin{pgfscope}%
\pgfsys@transformshift{3.530505in}{1.297360in}%
\pgfsys@useobject{currentmarker}{}%
\end{pgfscope}%
\begin{pgfscope}%
\pgfsys@transformshift{4.031515in}{1.573756in}%
\pgfsys@useobject{currentmarker}{}%
\end{pgfscope}%
\begin{pgfscope}%
\pgfsys@transformshift{4.532525in}{1.874182in}%
\pgfsys@useobject{currentmarker}{}%
\end{pgfscope}%
\begin{pgfscope}%
\pgfsys@transformshift{5.033535in}{2.193653in}%
\pgfsys@useobject{currentmarker}{}%
\end{pgfscope}%
\begin{pgfscope}%
\pgfsys@transformshift{5.534545in}{2.690247in}%
\pgfsys@useobject{currentmarker}{}%
\end{pgfscope}%
\end{pgfscope}%
\begin{pgfscope}%
\pgfpathrectangle{\pgfqpoint{0.800000in}{0.528000in}}{\pgfqpoint{4.960000in}{3.696000in}}%
\pgfusepath{clip}%
\pgfsetbuttcap%
\pgfsetroundjoin%
\definecolor{currentfill}{rgb}{0.501961,0.501961,0.501961}%
\pgfsetfillcolor{currentfill}%
\pgfsetlinewidth{1.003750pt}%
\definecolor{currentstroke}{rgb}{0.501961,0.501961,0.501961}%
\pgfsetstrokecolor{currentstroke}%
\pgfsetdash{}{0pt}%
\pgfsys@defobject{currentmarker}{\pgfqpoint{-0.020833in}{-0.020833in}}{\pgfqpoint{0.020833in}{0.020833in}}{%
\pgfpathmoveto{\pgfqpoint{0.000000in}{-0.020833in}}%
\pgfpathcurveto{\pgfqpoint{0.005525in}{-0.020833in}}{\pgfqpoint{0.010825in}{-0.018638in}}{\pgfqpoint{0.014731in}{-0.014731in}}%
\pgfpathcurveto{\pgfqpoint{0.018638in}{-0.010825in}}{\pgfqpoint{0.020833in}{-0.005525in}}{\pgfqpoint{0.020833in}{0.000000in}}%
\pgfpathcurveto{\pgfqpoint{0.020833in}{0.005525in}}{\pgfqpoint{0.018638in}{0.010825in}}{\pgfqpoint{0.014731in}{0.014731in}}%
\pgfpathcurveto{\pgfqpoint{0.010825in}{0.018638in}}{\pgfqpoint{0.005525in}{0.020833in}}{\pgfqpoint{0.000000in}{0.020833in}}%
\pgfpathcurveto{\pgfqpoint{-0.005525in}{0.020833in}}{\pgfqpoint{-0.010825in}{0.018638in}}{\pgfqpoint{-0.014731in}{0.014731in}}%
\pgfpathcurveto{\pgfqpoint{-0.018638in}{0.010825in}}{\pgfqpoint{-0.020833in}{0.005525in}}{\pgfqpoint{-0.020833in}{0.000000in}}%
\pgfpathcurveto{\pgfqpoint{-0.020833in}{-0.005525in}}{\pgfqpoint{-0.018638in}{-0.010825in}}{\pgfqpoint{-0.014731in}{-0.014731in}}%
\pgfpathcurveto{\pgfqpoint{-0.010825in}{-0.018638in}}{\pgfqpoint{-0.005525in}{-0.020833in}}{\pgfqpoint{0.000000in}{-0.020833in}}%
\pgfpathclose%
\pgfusepath{stroke,fill}%
}%
\begin{pgfscope}%
\pgfsys@transformshift{1.025455in}{0.968384in}%
\pgfsys@useobject{currentmarker}{}%
\end{pgfscope}%
\begin{pgfscope}%
\pgfsys@transformshift{1.526465in}{0.708754in}%
\pgfsys@useobject{currentmarker}{}%
\end{pgfscope}%
\begin{pgfscope}%
\pgfsys@transformshift{2.027475in}{0.722791in}%
\pgfsys@useobject{currentmarker}{}%
\end{pgfscope}%
\begin{pgfscope}%
\pgfsys@transformshift{2.528485in}{0.853820in}%
\pgfsys@useobject{currentmarker}{}%
\end{pgfscope}%
\begin{pgfscope}%
\pgfsys@transformshift{3.029495in}{1.052908in}%
\pgfsys@useobject{currentmarker}{}%
\end{pgfscope}%
\begin{pgfscope}%
\pgfsys@transformshift{3.530505in}{1.297070in}%
\pgfsys@useobject{currentmarker}{}%
\end{pgfscope}%
\begin{pgfscope}%
\pgfsys@transformshift{4.031515in}{1.573447in}%
\pgfsys@useobject{currentmarker}{}%
\end{pgfscope}%
\begin{pgfscope}%
\pgfsys@transformshift{4.532525in}{1.874119in}%
\pgfsys@useobject{currentmarker}{}%
\end{pgfscope}%
\begin{pgfscope}%
\pgfsys@transformshift{5.033535in}{2.196570in}%
\pgfsys@useobject{currentmarker}{}%
\end{pgfscope}%
\begin{pgfscope}%
\pgfsys@transformshift{5.534545in}{2.528300in}%
\pgfsys@useobject{currentmarker}{}%
\end{pgfscope}%
\end{pgfscope}%
\begin{pgfscope}%
\pgfsetrectcap%
\pgfsetmiterjoin%
\pgfsetlinewidth{0.803000pt}%
\definecolor{currentstroke}{rgb}{0.000000,0.000000,0.000000}%
\pgfsetstrokecolor{currentstroke}%
\pgfsetdash{}{0pt}%
\pgfpathmoveto{\pgfqpoint{0.800000in}{0.528000in}}%
\pgfpathlineto{\pgfqpoint{0.800000in}{4.224000in}}%
\pgfusepath{stroke}%
\end{pgfscope}%
\begin{pgfscope}%
\pgfsetrectcap%
\pgfsetmiterjoin%
\pgfsetlinewidth{0.803000pt}%
\definecolor{currentstroke}{rgb}{0.000000,0.000000,0.000000}%
\pgfsetstrokecolor{currentstroke}%
\pgfsetdash{}{0pt}%
\pgfpathmoveto{\pgfqpoint{5.760000in}{0.528000in}}%
\pgfpathlineto{\pgfqpoint{5.760000in}{4.224000in}}%
\pgfusepath{stroke}%
\end{pgfscope}%
\begin{pgfscope}%
\pgfsetrectcap%
\pgfsetmiterjoin%
\pgfsetlinewidth{0.803000pt}%
\definecolor{currentstroke}{rgb}{0.000000,0.000000,0.000000}%
\pgfsetstrokecolor{currentstroke}%
\pgfsetdash{}{0pt}%
\pgfpathmoveto{\pgfqpoint{0.800000in}{0.528000in}}%
\pgfpathlineto{\pgfqpoint{5.760000in}{0.528000in}}%
\pgfusepath{stroke}%
\end{pgfscope}%
\begin{pgfscope}%
\pgfsetrectcap%
\pgfsetmiterjoin%
\pgfsetlinewidth{0.803000pt}%
\definecolor{currentstroke}{rgb}{0.000000,0.000000,0.000000}%
\pgfsetstrokecolor{currentstroke}%
\pgfsetdash{}{0pt}%
\pgfpathmoveto{\pgfqpoint{0.800000in}{4.224000in}}%
\pgfpathlineto{\pgfqpoint{5.760000in}{4.224000in}}%
\pgfusepath{stroke}%
\end{pgfscope}%
\begin{pgfscope}%
\pgfsetbuttcap%
\pgfsetmiterjoin%
\definecolor{currentfill}{rgb}{1.000000,1.000000,1.000000}%
\pgfsetfillcolor{currentfill}%
\pgfsetfillopacity{0.800000}%
\pgfsetlinewidth{1.003750pt}%
\definecolor{currentstroke}{rgb}{0.800000,0.800000,0.800000}%
\pgfsetstrokecolor{currentstroke}%
\pgfsetstrokeopacity{0.800000}%
\pgfsetdash{}{0pt}%
\pgfpathmoveto{\pgfqpoint{4.638358in}{2.887779in}}%
\pgfpathlineto{\pgfqpoint{5.662778in}{2.887779in}}%
\pgfpathquadraticcurveto{\pgfqpoint{5.690556in}{2.887779in}}{\pgfqpoint{5.690556in}{2.915556in}}%
\pgfpathlineto{\pgfqpoint{5.690556in}{4.126778in}}%
\pgfpathquadraticcurveto{\pgfqpoint{5.690556in}{4.154556in}}{\pgfqpoint{5.662778in}{4.154556in}}%
\pgfpathlineto{\pgfqpoint{4.638358in}{4.154556in}}%
\pgfpathquadraticcurveto{\pgfqpoint{4.610580in}{4.154556in}}{\pgfqpoint{4.610580in}{4.126778in}}%
\pgfpathlineto{\pgfqpoint{4.610580in}{2.915556in}}%
\pgfpathquadraticcurveto{\pgfqpoint{4.610580in}{2.887779in}}{\pgfqpoint{4.638358in}{2.887779in}}%
\pgfpathclose%
\pgfusepath{stroke,fill}%
\end{pgfscope}%
\begin{pgfscope}%
\pgfsetbuttcap%
\pgfsetroundjoin%
\definecolor{currentfill}{rgb}{1.000000,0.000000,0.000000}%
\pgfsetfillcolor{currentfill}%
\pgfsetlinewidth{1.003750pt}%
\definecolor{currentstroke}{rgb}{1.000000,0.000000,0.000000}%
\pgfsetstrokecolor{currentstroke}%
\pgfsetdash{}{0pt}%
\pgfsys@defobject{currentmarker}{\pgfqpoint{-0.020833in}{-0.020833in}}{\pgfqpoint{0.020833in}{0.020833in}}{%
\pgfpathmoveto{\pgfqpoint{0.000000in}{-0.020833in}}%
\pgfpathcurveto{\pgfqpoint{0.005525in}{-0.020833in}}{\pgfqpoint{0.010825in}{-0.018638in}}{\pgfqpoint{0.014731in}{-0.014731in}}%
\pgfpathcurveto{\pgfqpoint{0.018638in}{-0.010825in}}{\pgfqpoint{0.020833in}{-0.005525in}}{\pgfqpoint{0.020833in}{0.000000in}}%
\pgfpathcurveto{\pgfqpoint{0.020833in}{0.005525in}}{\pgfqpoint{0.018638in}{0.010825in}}{\pgfqpoint{0.014731in}{0.014731in}}%
\pgfpathcurveto{\pgfqpoint{0.010825in}{0.018638in}}{\pgfqpoint{0.005525in}{0.020833in}}{\pgfqpoint{0.000000in}{0.020833in}}%
\pgfpathcurveto{\pgfqpoint{-0.005525in}{0.020833in}}{\pgfqpoint{-0.010825in}{0.018638in}}{\pgfqpoint{-0.014731in}{0.014731in}}%
\pgfpathcurveto{\pgfqpoint{-0.018638in}{0.010825in}}{\pgfqpoint{-0.020833in}{0.005525in}}{\pgfqpoint{-0.020833in}{0.000000in}}%
\pgfpathcurveto{\pgfqpoint{-0.020833in}{-0.005525in}}{\pgfqpoint{-0.018638in}{-0.010825in}}{\pgfqpoint{-0.014731in}{-0.014731in}}%
\pgfpathcurveto{\pgfqpoint{-0.010825in}{-0.018638in}}{\pgfqpoint{-0.005525in}{-0.020833in}}{\pgfqpoint{0.000000in}{-0.020833in}}%
\pgfpathclose%
\pgfusepath{stroke,fill}%
}%
\begin{pgfscope}%
\pgfsys@transformshift{4.805024in}{4.042088in}%
\pgfsys@useobject{currentmarker}{}%
\end{pgfscope}%
\end{pgfscope}%
\begin{pgfscope}%
\definecolor{textcolor}{rgb}{0.000000,0.000000,0.000000}%
\pgfsetstrokecolor{textcolor}%
\pgfsetfillcolor{textcolor}%
\pgftext[x=5.055024in,y=3.993477in,left,base]{\color{textcolor}\sffamily\fontsize{10.000000}{12.000000}\selectfont \(\displaystyle \mathcal{R}_1 = \mathbf{1}\)}%
\end{pgfscope}%
\begin{pgfscope}%
\pgfsetbuttcap%
\pgfsetroundjoin%
\definecolor{currentfill}{rgb}{1.000000,0.647059,0.000000}%
\pgfsetfillcolor{currentfill}%
\pgfsetlinewidth{1.003750pt}%
\definecolor{currentstroke}{rgb}{1.000000,0.647059,0.000000}%
\pgfsetstrokecolor{currentstroke}%
\pgfsetdash{}{0pt}%
\pgfsys@defobject{currentmarker}{\pgfqpoint{-0.020833in}{-0.020833in}}{\pgfqpoint{0.020833in}{0.020833in}}{%
\pgfpathmoveto{\pgfqpoint{0.000000in}{-0.020833in}}%
\pgfpathcurveto{\pgfqpoint{0.005525in}{-0.020833in}}{\pgfqpoint{0.010825in}{-0.018638in}}{\pgfqpoint{0.014731in}{-0.014731in}}%
\pgfpathcurveto{\pgfqpoint{0.018638in}{-0.010825in}}{\pgfqpoint{0.020833in}{-0.005525in}}{\pgfqpoint{0.020833in}{0.000000in}}%
\pgfpathcurveto{\pgfqpoint{0.020833in}{0.005525in}}{\pgfqpoint{0.018638in}{0.010825in}}{\pgfqpoint{0.014731in}{0.014731in}}%
\pgfpathcurveto{\pgfqpoint{0.010825in}{0.018638in}}{\pgfqpoint{0.005525in}{0.020833in}}{\pgfqpoint{0.000000in}{0.020833in}}%
\pgfpathcurveto{\pgfqpoint{-0.005525in}{0.020833in}}{\pgfqpoint{-0.010825in}{0.018638in}}{\pgfqpoint{-0.014731in}{0.014731in}}%
\pgfpathcurveto{\pgfqpoint{-0.018638in}{0.010825in}}{\pgfqpoint{-0.020833in}{0.005525in}}{\pgfqpoint{-0.020833in}{0.000000in}}%
\pgfpathcurveto{\pgfqpoint{-0.020833in}{-0.005525in}}{\pgfqpoint{-0.018638in}{-0.010825in}}{\pgfqpoint{-0.014731in}{-0.014731in}}%
\pgfpathcurveto{\pgfqpoint{-0.010825in}{-0.018638in}}{\pgfqpoint{-0.005525in}{-0.020833in}}{\pgfqpoint{0.000000in}{-0.020833in}}%
\pgfpathclose%
\pgfusepath{stroke,fill}%
}%
\begin{pgfscope}%
\pgfsys@transformshift{4.805024in}{3.838231in}%
\pgfsys@useobject{currentmarker}{}%
\end{pgfscope}%
\end{pgfscope}%
\begin{pgfscope}%
\definecolor{textcolor}{rgb}{0.000000,0.000000,0.000000}%
\pgfsetstrokecolor{textcolor}%
\pgfsetfillcolor{textcolor}%
\pgftext[x=5.055024in,y=3.789620in,left,base]{\color{textcolor}\sffamily\fontsize{10.000000}{12.000000}\selectfont \(\displaystyle \mathcal{R}_2\)}%
\end{pgfscope}%
\begin{pgfscope}%
\pgfsetbuttcap%
\pgfsetroundjoin%
\definecolor{currentfill}{rgb}{0.000000,0.501961,0.000000}%
\pgfsetfillcolor{currentfill}%
\pgfsetlinewidth{1.003750pt}%
\definecolor{currentstroke}{rgb}{0.000000,0.501961,0.000000}%
\pgfsetstrokecolor{currentstroke}%
\pgfsetdash{}{0pt}%
\pgfsys@defobject{currentmarker}{\pgfqpoint{-0.020833in}{-0.020833in}}{\pgfqpoint{0.020833in}{0.020833in}}{%
\pgfpathmoveto{\pgfqpoint{0.000000in}{-0.020833in}}%
\pgfpathcurveto{\pgfqpoint{0.005525in}{-0.020833in}}{\pgfqpoint{0.010825in}{-0.018638in}}{\pgfqpoint{0.014731in}{-0.014731in}}%
\pgfpathcurveto{\pgfqpoint{0.018638in}{-0.010825in}}{\pgfqpoint{0.020833in}{-0.005525in}}{\pgfqpoint{0.020833in}{0.000000in}}%
\pgfpathcurveto{\pgfqpoint{0.020833in}{0.005525in}}{\pgfqpoint{0.018638in}{0.010825in}}{\pgfqpoint{0.014731in}{0.014731in}}%
\pgfpathcurveto{\pgfqpoint{0.010825in}{0.018638in}}{\pgfqpoint{0.005525in}{0.020833in}}{\pgfqpoint{0.000000in}{0.020833in}}%
\pgfpathcurveto{\pgfqpoint{-0.005525in}{0.020833in}}{\pgfqpoint{-0.010825in}{0.018638in}}{\pgfqpoint{-0.014731in}{0.014731in}}%
\pgfpathcurveto{\pgfqpoint{-0.018638in}{0.010825in}}{\pgfqpoint{-0.020833in}{0.005525in}}{\pgfqpoint{-0.020833in}{0.000000in}}%
\pgfpathcurveto{\pgfqpoint{-0.020833in}{-0.005525in}}{\pgfqpoint{-0.018638in}{-0.010825in}}{\pgfqpoint{-0.014731in}{-0.014731in}}%
\pgfpathcurveto{\pgfqpoint{-0.010825in}{-0.018638in}}{\pgfqpoint{-0.005525in}{-0.020833in}}{\pgfqpoint{0.000000in}{-0.020833in}}%
\pgfpathclose%
\pgfusepath{stroke,fill}%
}%
\begin{pgfscope}%
\pgfsys@transformshift{4.805024in}{3.634374in}%
\pgfsys@useobject{currentmarker}{}%
\end{pgfscope}%
\end{pgfscope}%
\begin{pgfscope}%
\definecolor{textcolor}{rgb}{0.000000,0.000000,0.000000}%
\pgfsetstrokecolor{textcolor}%
\pgfsetfillcolor{textcolor}%
\pgftext[x=5.055024in,y=3.585762in,left,base]{\color{textcolor}\sffamily\fontsize{10.000000}{12.000000}\selectfont \(\displaystyle \mathcal{R}_3\)}%
\end{pgfscope}%
\begin{pgfscope}%
\pgfsetbuttcap%
\pgfsetroundjoin%
\definecolor{currentfill}{rgb}{0.603922,0.803922,0.196078}%
\pgfsetfillcolor{currentfill}%
\pgfsetlinewidth{1.003750pt}%
\definecolor{currentstroke}{rgb}{0.603922,0.803922,0.196078}%
\pgfsetstrokecolor{currentstroke}%
\pgfsetdash{}{0pt}%
\pgfsys@defobject{currentmarker}{\pgfqpoint{-0.020833in}{-0.020833in}}{\pgfqpoint{0.020833in}{0.020833in}}{%
\pgfpathmoveto{\pgfqpoint{0.000000in}{-0.020833in}}%
\pgfpathcurveto{\pgfqpoint{0.005525in}{-0.020833in}}{\pgfqpoint{0.010825in}{-0.018638in}}{\pgfqpoint{0.014731in}{-0.014731in}}%
\pgfpathcurveto{\pgfqpoint{0.018638in}{-0.010825in}}{\pgfqpoint{0.020833in}{-0.005525in}}{\pgfqpoint{0.020833in}{0.000000in}}%
\pgfpathcurveto{\pgfqpoint{0.020833in}{0.005525in}}{\pgfqpoint{0.018638in}{0.010825in}}{\pgfqpoint{0.014731in}{0.014731in}}%
\pgfpathcurveto{\pgfqpoint{0.010825in}{0.018638in}}{\pgfqpoint{0.005525in}{0.020833in}}{\pgfqpoint{0.000000in}{0.020833in}}%
\pgfpathcurveto{\pgfqpoint{-0.005525in}{0.020833in}}{\pgfqpoint{-0.010825in}{0.018638in}}{\pgfqpoint{-0.014731in}{0.014731in}}%
\pgfpathcurveto{\pgfqpoint{-0.018638in}{0.010825in}}{\pgfqpoint{-0.020833in}{0.005525in}}{\pgfqpoint{-0.020833in}{0.000000in}}%
\pgfpathcurveto{\pgfqpoint{-0.020833in}{-0.005525in}}{\pgfqpoint{-0.018638in}{-0.010825in}}{\pgfqpoint{-0.014731in}{-0.014731in}}%
\pgfpathcurveto{\pgfqpoint{-0.010825in}{-0.018638in}}{\pgfqpoint{-0.005525in}{-0.020833in}}{\pgfqpoint{0.000000in}{-0.020833in}}%
\pgfpathclose%
\pgfusepath{stroke,fill}%
}%
\begin{pgfscope}%
\pgfsys@transformshift{4.805024in}{3.430516in}%
\pgfsys@useobject{currentmarker}{}%
\end{pgfscope}%
\end{pgfscope}%
\begin{pgfscope}%
\definecolor{textcolor}{rgb}{0.000000,0.000000,0.000000}%
\pgfsetstrokecolor{textcolor}%
\pgfsetfillcolor{textcolor}%
\pgftext[x=5.055024in,y=3.381905in,left,base]{\color{textcolor}\sffamily\fontsize{10.000000}{12.000000}\selectfont \(\displaystyle \mathcal{R}_4\)}%
\end{pgfscope}%
\begin{pgfscope}%
\pgfsetbuttcap%
\pgfsetroundjoin%
\definecolor{currentfill}{rgb}{0.000000,0.000000,1.000000}%
\pgfsetfillcolor{currentfill}%
\pgfsetlinewidth{1.003750pt}%
\definecolor{currentstroke}{rgb}{0.000000,0.000000,1.000000}%
\pgfsetstrokecolor{currentstroke}%
\pgfsetdash{}{0pt}%
\pgfsys@defobject{currentmarker}{\pgfqpoint{-0.020833in}{-0.020833in}}{\pgfqpoint{0.020833in}{0.020833in}}{%
\pgfpathmoveto{\pgfqpoint{0.000000in}{-0.020833in}}%
\pgfpathcurveto{\pgfqpoint{0.005525in}{-0.020833in}}{\pgfqpoint{0.010825in}{-0.018638in}}{\pgfqpoint{0.014731in}{-0.014731in}}%
\pgfpathcurveto{\pgfqpoint{0.018638in}{-0.010825in}}{\pgfqpoint{0.020833in}{-0.005525in}}{\pgfqpoint{0.020833in}{0.000000in}}%
\pgfpathcurveto{\pgfqpoint{0.020833in}{0.005525in}}{\pgfqpoint{0.018638in}{0.010825in}}{\pgfqpoint{0.014731in}{0.014731in}}%
\pgfpathcurveto{\pgfqpoint{0.010825in}{0.018638in}}{\pgfqpoint{0.005525in}{0.020833in}}{\pgfqpoint{0.000000in}{0.020833in}}%
\pgfpathcurveto{\pgfqpoint{-0.005525in}{0.020833in}}{\pgfqpoint{-0.010825in}{0.018638in}}{\pgfqpoint{-0.014731in}{0.014731in}}%
\pgfpathcurveto{\pgfqpoint{-0.018638in}{0.010825in}}{\pgfqpoint{-0.020833in}{0.005525in}}{\pgfqpoint{-0.020833in}{0.000000in}}%
\pgfpathcurveto{\pgfqpoint{-0.020833in}{-0.005525in}}{\pgfqpoint{-0.018638in}{-0.010825in}}{\pgfqpoint{-0.014731in}{-0.014731in}}%
\pgfpathcurveto{\pgfqpoint{-0.010825in}{-0.018638in}}{\pgfqpoint{-0.005525in}{-0.020833in}}{\pgfqpoint{0.000000in}{-0.020833in}}%
\pgfpathclose%
\pgfusepath{stroke,fill}%
}%
\begin{pgfscope}%
\pgfsys@transformshift{4.805024in}{3.226659in}%
\pgfsys@useobject{currentmarker}{}%
\end{pgfscope}%
\end{pgfscope}%
\begin{pgfscope}%
\definecolor{textcolor}{rgb}{0.000000,0.000000,0.000000}%
\pgfsetstrokecolor{textcolor}%
\pgfsetfillcolor{textcolor}%
\pgftext[x=5.055024in,y=3.178048in,left,base]{\color{textcolor}\sffamily\fontsize{10.000000}{12.000000}\selectfont \(\displaystyle \mathcal{R}_5 = \mathbf{adj}\)}%
\end{pgfscope}%
\begin{pgfscope}%
\pgfsetbuttcap%
\pgfsetroundjoin%
\definecolor{currentfill}{rgb}{0.501961,0.501961,0.501961}%
\pgfsetfillcolor{currentfill}%
\pgfsetlinewidth{1.003750pt}%
\definecolor{currentstroke}{rgb}{0.501961,0.501961,0.501961}%
\pgfsetstrokecolor{currentstroke}%
\pgfsetdash{}{0pt}%
\pgfsys@defobject{currentmarker}{\pgfqpoint{-0.020833in}{-0.020833in}}{\pgfqpoint{0.020833in}{0.020833in}}{%
\pgfpathmoveto{\pgfqpoint{0.000000in}{-0.020833in}}%
\pgfpathcurveto{\pgfqpoint{0.005525in}{-0.020833in}}{\pgfqpoint{0.010825in}{-0.018638in}}{\pgfqpoint{0.014731in}{-0.014731in}}%
\pgfpathcurveto{\pgfqpoint{0.018638in}{-0.010825in}}{\pgfqpoint{0.020833in}{-0.005525in}}{\pgfqpoint{0.020833in}{0.000000in}}%
\pgfpathcurveto{\pgfqpoint{0.020833in}{0.005525in}}{\pgfqpoint{0.018638in}{0.010825in}}{\pgfqpoint{0.014731in}{0.014731in}}%
\pgfpathcurveto{\pgfqpoint{0.010825in}{0.018638in}}{\pgfqpoint{0.005525in}{0.020833in}}{\pgfqpoint{0.000000in}{0.020833in}}%
\pgfpathcurveto{\pgfqpoint{-0.005525in}{0.020833in}}{\pgfqpoint{-0.010825in}{0.018638in}}{\pgfqpoint{-0.014731in}{0.014731in}}%
\pgfpathcurveto{\pgfqpoint{-0.018638in}{0.010825in}}{\pgfqpoint{-0.020833in}{0.005525in}}{\pgfqpoint{-0.020833in}{0.000000in}}%
\pgfpathcurveto{\pgfqpoint{-0.020833in}{-0.005525in}}{\pgfqpoint{-0.018638in}{-0.010825in}}{\pgfqpoint{-0.014731in}{-0.014731in}}%
\pgfpathcurveto{\pgfqpoint{-0.010825in}{-0.018638in}}{\pgfqpoint{-0.005525in}{-0.020833in}}{\pgfqpoint{0.000000in}{-0.020833in}}%
\pgfpathclose%
\pgfusepath{stroke,fill}%
}%
\begin{pgfscope}%
\pgfsys@transformshift{4.805024in}{3.020835in}%
\pgfsys@useobject{currentmarker}{}%
\end{pgfscope}%
\end{pgfscope}%
\begin{pgfscope}%
\definecolor{textcolor}{rgb}{0.000000,0.000000,0.000000}%
\pgfsetstrokecolor{textcolor}%
\pgfsetfillcolor{textcolor}%
\pgftext[x=5.055024in,y=2.972224in,left,base]{\color{textcolor}\sffamily\fontsize{10.000000}{12.000000}\selectfont \(\displaystyle \mathcal{R}_6\)}%
\end{pgfscope}%
\end{pgfpicture}%
\makeatother%
\endgroup%